# Optimal Estimation of States in Quantum Image Processing

**Mario Mastriani**

DLQS LLC, 4431 NW 63RD Drive, Coconut Creek, FL 33073, USA.
mmastri@gmail.com

*Abstract*—An optimal estimator of quantum states based on a modified Kalman's Filter is presented in this work. Such estimator acts after state measurement, allowing to obtain an optimal estimation of quantum state resulting in the output of any quantum image algorithm**.** Besides, a new criteria, logic, and arithmetic based on projections onto vertical axis of Bloch's Sphere exclusively are presented too. This approach will allow us: 1) a simpler development of logic and arithmetic quantum operations, where they will closer to those used in the classical digital image processing algorithms, 2) building simple and robust classical-to-quantum and quantum-to-classical  interfaces. Said so far is extended to quantum algorithms outside image processing too. In a special section on metrics and simulations, three new metrics based on the comparison between the classical and quantum versions algorithms for filtering and edge detection of images are presented. Notable differences between the results of classical and quantum versions of such algorithms (outside and inside of quantum computer, respectively) show the need for modeling state and measurement noise inside estimation scheme.

## 1 Introduction

Quantum computation and quantum information is the study of the information processing tasks that can be accomplished using quantum mechanical systems. Like many simple but profound ideas it was a long time before anybody thought of doing information processing using quantum mechanical systems [1].

To see why this is the case, we must go back in time and look in turn at each of the fields which have contributed fundamental ideas to quantum computation and quantum information – quantum mechanics, computer science, information theory, and cryptography. As we take our short historical tour of these fields, think of yourself first as a physicist, then as a computer scientist, then as an information theorist, and finally as a cryptographer, in order to get some feel for the disparate perspectives which have come together in quantum computation and quantum information.

Today's computers—both in theory (Turing's machines) and practice (PCs)—are based on classical physics. However, modern quantum physics tells us that the world behaves quite differently. A quantum system can be in a superposition of many different states at the same time, and can exhibit interference effects during the course of its evolution. Moreover, spatially separated quantum systems may be entangled with each other and operations may have "non-local" effects because of this.

Quantum computation is the field that investigates the computational power and other properties of computers based on quantum-mechanical principles. An important objective is to find quantum algorithms that are significantly faster than any classical algorithm solving the same problem. The field started in the early 1980s with suggestions for analog quantum computers by Paul Benioff [2] and Richard Feynman [3, 4], and reached more digital ground when in 1985 David Deutsch defined the universal quantum Turing machine [5]. The following years saw only sparse activity, notably the development of the first algorithms by

Deutsch and Jozsa [6] and by Simon [7], and the development of quantum complexity theory by Bernstein and Vazirani [8]. However, interest in the field increased tremendously after Peter Shor's very surprising discovery of efficient quantum algorithms (or simulations on a quantum computer) for the problems of integer factorization and discrete logarithms in 1994 [9]. Since most of current classical cryptography is based on the assumption that these two problems are computationally hard, the ability to actually build and use a quantum computer would allow us to break most current classical cryptographic systems, notably the RSA system [10, 11]. In contrast, a quantum form of cryptography due to Bennett and Brassard [12] is unbreakable even for quantum computers.

What do we mean by 'efficient' versus 'inefficient' simulations of a quantum computer? Many of key notions needed to answer this question were actually invented before the notion of a quantum computer had even arisen. In particular, the idea of efficient and inefficient algorithms was made mathematically precise by the field of computational complexity. Roughly speaking, an efficient algorithm which runs in polynomial time is the size of the problem solved. In contrast, an inefficient algorithm requires super polynomial (typically exponential) time. What was noticed in the late 1960s and early 1970s was that it seemed as though the Turing machine model of computation was at least as powerful as any other model of computation, in the sense that a problem which could be solved efficiently in some model of computation could also be solved efficiently in the Turing machine model, by using the Turing machine to simulate the other model of computation. This observation was codified into a strengthened version of the Church–Turing thesis:

*Any algorithmic process can be simulated efficiently using a Turing machine.*

The key strengthening in the strong Church–Turing thesis is the word efficiently. If the strong Church–Turing thesis is correct, then it implies that no matter what type of machine we use to perform our algorithms, that machine can be simulated efficiently using a standard Turing machine. This is an important strengthening, as it implies that for the purposes of analyzing whether a given computational task can be accomplished efficiently, we may restrict ourselves to the analysis of the Turing machine model of computation.

Randomized algorithms pose a challenge to the strong Church–Turing thesis, suggesting that there are efficiently soluble problems which, nevertheless, cannot be efficiently solved on a deterministic Turing machine. This challenge appears to be easily resolved by a simple modification of the strong Church–Turing thesis:

*Any algorithmic process can be simulated efficiently using a probabilistic Turing machine.*

This ad hoc modification of the strong Church–Turing thesis should leave we feeling rather queasy.

On the other hand, and as well as Hirota *et al* said inside the Introduction of their work [13]:

*Quantum computation has appeared in various areas of computer science such as information theory, cryptography, image processing, etc. [1] because there are inefficient tasks on classical computers that can be overcome by exploiting the power of the quantum computation. Processing and analysis of images in particular and visual information in general on classical computers have been studied extensively [14-17]. On quantum computers, the research on images has faced fundamental difficulties because the field is still in its infancy. To start with, what are quantum images or how do we represent images on quantum computers? Secondly, what should we do to prepare and process the quantum images on quantum computers?*

Precisely, these two questions represent the essence on which this paper is based, i.e., the correct (and more efficient) internal representation of an image in a quantum context, and its recovery, once processed internally. Thus, we recognize only 3 milestones in the brief history of quantum image processing, namely:

- all starts with the pioneering work of Prof. Salvador E. Venegas-Andraca [18-21] at Keble College, Oxford University, UK (currently at Tecnológico de Monterrey, Campus Estado de México), where he proposes quantum image representations such as Qubit Lattice [22], in fact, this is the first doctoral thesis in the specialty,

- the history continues with the quantum image representation via the Real Ket [23] of Prof. Jose I. Latorre Sentís, at Universitat de Barcelona, Spain, with a special interest on image compression in a quantum context, and finally,

- we arrive at the proposal of Prof. Kaoru Hirota *et al* [13] from Tokyo Institute of Technology, for a flexible representation of quantum images to provide a representation for images on quantum computers in the form of a normalized state which captures information about colors and their corresponding positions in the images.

These works marked the path and viability of quantum image processing, however, we believe that a new type of internal representation of images, which enables an easier representation of traditional Digital Image Processing algorithms in a quantum computer, as well as more easy and efficient recovery of images processed outside the quantum computer is imperative. Besides, we present a novel proposal to recover quantum state to the output of a quantum algorithm after its measurement via a modified Kalman's Filter [24-28], and Recursive Least Squares (RLS) filter [29-31], too. This is the essence of this work, which is organized as follows:

The basic principles of Quantum Information Processing are outlined in Section 2. Implementation Problems in Quantum Image Processing are presented in Section 3. The new approach for internal image representation is outlined in Section 4. In Section 5, we present the development of modified Kalman's filter for the optimal quantum state estimation. In Section 6, we show the proposed new interfaces classical-to-quantum and quantum-to-classical arising from the tools mentioned in the previous sections. In Section 7, we present a bit of Digital Image Processing. In Section 8, we discuss the more appropriate metrics for denoising and edge detection in a set of experimental results. Finally, Section 9 provides a conclusion and future works proposal of the paper.

## 2 Quantum Information Processing

In this section, we present the main concepts related to Quantum Information Processing, that is to say: qubit, Bloch's Sphere, Hilbert's Space, Schrödinger Equation, Unitary Operators, Quantum Circuits/Gates, and Quantum Algorithms.

2.1 Quantum bits (qubits) and Bloch's sphere

The bit is the fundamental concept of classical computation and classical information. Quantum computation and quantum information are built upon an analogous concept, the quantum bit, or qubit for short. In this section we introduce the properties of single and multiple qubits, comparing and contrasting their properties to those of classical bits [1].

The difference between bits and qubits is that a qubit can be in a state other than $|0\rangle$ or $|1\rangle$ [32, 33]. It is also possible to form linear combinations of states, often called superpositions:

$$|\psi\rangle = \alpha|0\rangle + \beta|1\rangle, \tag{1}$$

where $|\alpha|^2 + |\beta|^2 = 1$, with the states $|\alpha\rangle$ and $|\beta\rangle$ are understood as different polarization states of light. The numbers α and β are complex numbers, although for many purposes not much is lost by thinking of them as real numbers. In other words, the state of a qubit is a vector in a two-dimensional complex vector space. The special states $|0\rangle$ and $|1\rangle$ are known as computational basis states, and form an orthonormal basis for this vector space, being

$$|0\rangle = \begin{bmatrix} 1 \\ 0 \end{bmatrix}$$

and

$$|1\rangle = \begin{bmatrix} 0 \\ 1 \end{bmatrix}$$

One picture useful in thinking about qubits is the following geometric representation.

Because $|\alpha|^2 + |\beta|^2 = 1$, we may rewrite Equation (1) as

$$|\psi\rangle = e^{i\gamma}\left(cos\frac{\theta}{2}|0\rangle + e^{i\phi} sin\frac{\theta}{2}|1\rangle\right) = e^{i\gamma}\left(cos\frac{\theta}{2}|0\rangle + (cos\phi + i\, sin\phi) sin\frac{\theta}{2}|1\rangle\right) \quad (2)$$

where $0 \leq \theta \leq \pi$, $0 \leq \phi < 2\pi$. We can ignore the factor of $e^{i\gamma}$ out the front, because it has no observable effects [1], and for that reason we can effectively write

$$|\psi\rangle = cos\frac{\theta}{2}|0\rangle + e^{i\phi} sin\frac{\theta}{2}|1\rangle \quad (3)$$

The numbers $\theta$ and $\phi$ define a point on the unit three-dimensional sphere, as shown in Fig. 1.

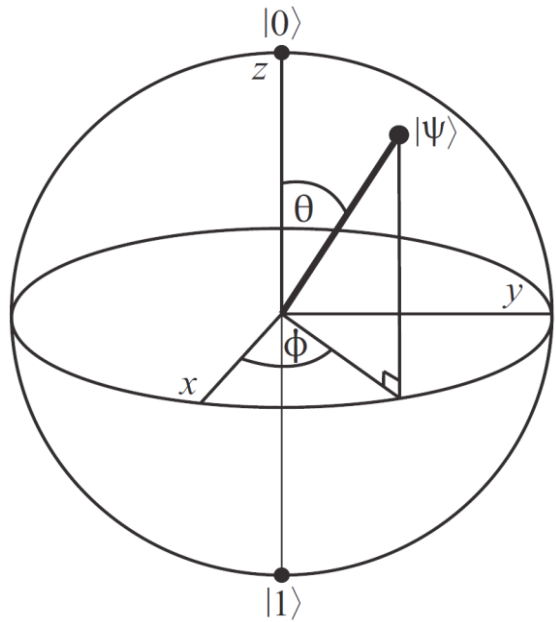

**Fig. 1** Bloch's Sphere.

Quantum mechanics is mathematically formulated in Hilbert space or projective Hilbert space. The space of pure states of a quantum system is given by the one-dimensional subspaces of the corresponding Hilbert space (or the "points" of the projective Hilbert space). In a two-dimensional Hilbert space this is simply the complex projective line, which is a geometrical sphere.

This sphere is often called the Bloch's sphere; it provides a useful means of visualizing the state of a single qubit, and often serves as an excellent testbed for ideas about quantum computation and quantum information. Many of the operations on single qubits which we describe later in this chapter are neatly described within the Bloch's sphere picture. However, it must be kept in mind that this intuition is limited because there is no a simple generalization of the Bloch's sphere known for multiple qubits [1, 32, 33].

Except in the case where $|\psi\rangle$ is one of the ket vectors $|0\rangle$ or $|1\rangle$ the representation is unique. The parameters $\theta$ and $\phi$, re-interpreted as spherical coordinates, specify a point $\vec{a} = (sin\theta cos\phi + sin\theta sin\phi + cos\theta)$ on the unit sphere in $\mathbb{R}^3$ (according to Eq. 2).

In the special case that $\phi = 0°$, it is easier to observe the corresponding $\alpha$ and $\beta$ projections, see Fig. 2.

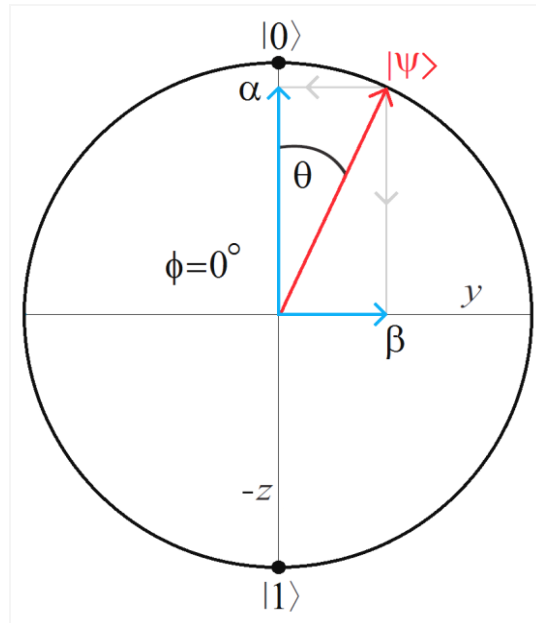

**Fig. 2** $\alpha$ and $\beta$ projections for $\phi = 0°$.

Fig. 3 highlights all components (details) concerning the Bloch's sphere, namely

*Spin down* $= |\downarrow\rangle = |0\rangle = \begin{bmatrix} 1 \\ 0 \end{bmatrix} =$ *qubit basis state = North Pole*

and

*Spin up* $= |\uparrow\rangle = |1\rangle = \begin{bmatrix} 0 \\ 1 \end{bmatrix} =$ *qubit basis state = South Pole*

Both poles play a fundamental role in the development of the subsequent sections. Besides, a very important concept to the affections of the development of this work, i.e., the notion of latitude (parallel) on the Bloch's sphere is hinted. Such parallel as shown in green in Fig. 3. However, it is on Fig. 4 where we can see the complete coexistence of poles, parallels and meridians on the sphere, including computational basis states ($|0\rangle, |1\rangle$). The poles and the parallels form the geometric bases of criteria and logic needed to implement our algorithms for quantum image processing and the classical-to-quantum, and quantum-to-classical interfaces.

2.2 Schrödinger´s equation and unitary operators

A quantum state can be transformed into another state by a unitary operator, symbolized as $U$, with $U^\dagger U = I$ (where $I$ is the identity matrix), which is required to preserve inner products: If we transform $|\chi\rangle$ and $|\psi\rangle$ to $U|\chi\rangle$ and $U|\psi\rangle$, then $\langle\chi|UU|\psi\rangle = \langle\chi|\psi\rangle$. In particular, unitary operators preserve lengths: $\langle\psi|UU|\psi\rangle = \langle\psi|\psi\rangle = 1$.

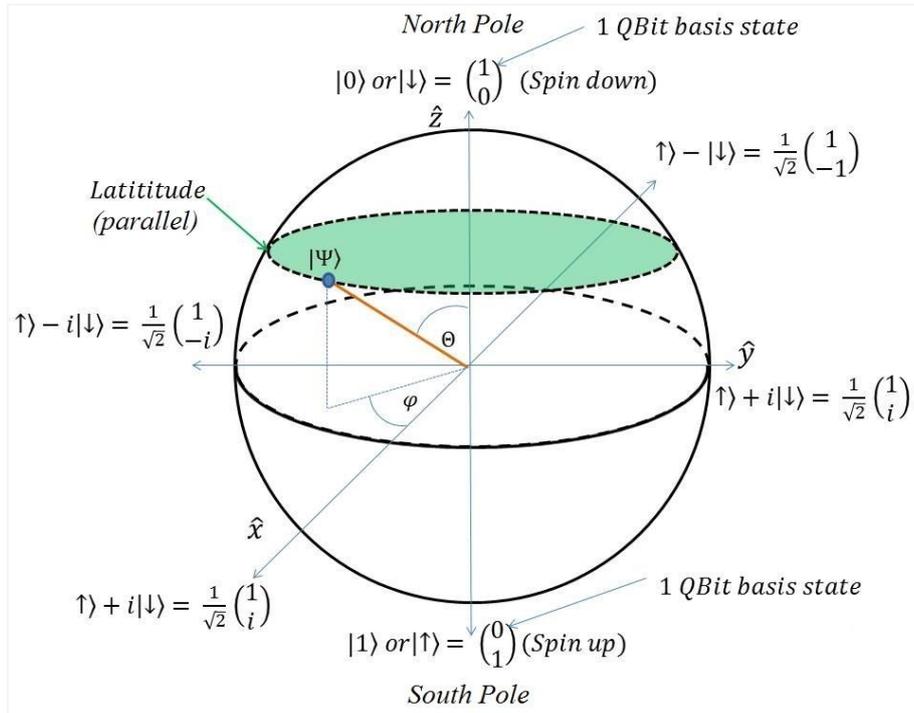

**Fig. 3** Details of the poles, as well as an example of parallel and several qubit states on the sphere.

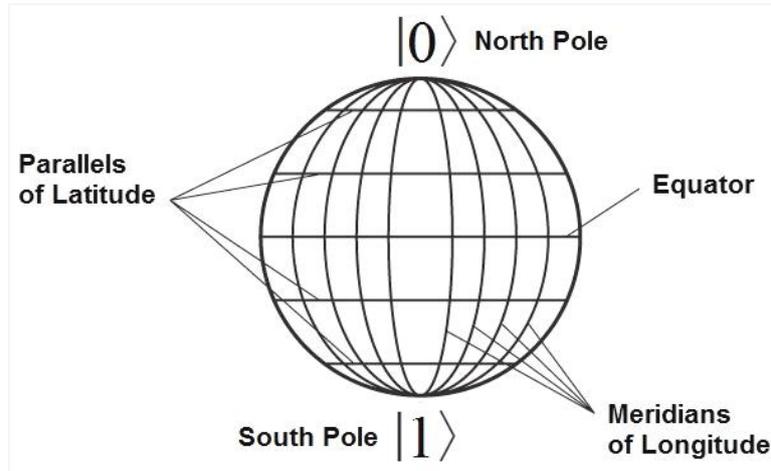

**Fig. 4** Coexistence of poles, parallels and meridians on the sphere.

On the other hand, the unitary operator satisfies the following differential equation known as the Schrödinger equation [1, 32-34]:

$$\frac{d}{dt}U(t) = \frac{-i\hat{H}}{\hbar}U(t) \tag{4}$$

where $\hat{H}$ represents the Hamiltonian matrix of the Schrödinger equation, $i = \sqrt[2]{-1}$, and $\hbar$ is the Planck constant. Multiplying both sides of Eq.(4) by $|\psi(0)\rangle$ and setting $|\psi(t)\rangle = U(t)|\psi(0)\rangle$ yields

$$\frac{d}{dt}|\psi(t)\rangle = \frac{-i\hat{H}}{\hbar}|\psi(t)\rangle \tag{5}$$

The solution to the Schrödinger equation is given by the matrix exponential of the Hamiltonian matrix:

$$U(t) = e^{\frac{-i\hat{H}t}{\hbar}} \quad (6)$$

Thus the probability amplitudes evolve across time according to the following equation:

$$|\psi(t)\rangle = e^{\frac{-i\hat{H}t}{\hbar}} |\psi(0)\rangle \quad (7)$$

Equation 7 is the main piece in building circuits, gates and quantum algorithms, being $U$ who represents such elements [1].

Finally, the discrete version of Eq.(5) is

$$|\psi_{t+1}\rangle = \frac{-i\hat{H}}{\hbar} |\psi_t\rangle \quad (8)$$

Equation 8 is the foundation on which we build the optimal estimator of quantum states.

2.3 Quantum Circuits, Gates and Algorithms

As we can see in Fig. 5, and remember Eq.(8), the quantum algorithm (identical case to circuits and gates) viewed as a transfer (or mapping input-to-output) has two types of output:

*a)* the result of algorithm (circuit of gate), i.e., $|\psi_{t+1}\rangle$

b) part of the input $|\psi_t\rangle$, i.e., $|\underline{\psi}_t\rangle$ (underlined $|\psi_t\rangle$), in order to impart reversibility to the circuit, which is a critical need in quantum computing [1].

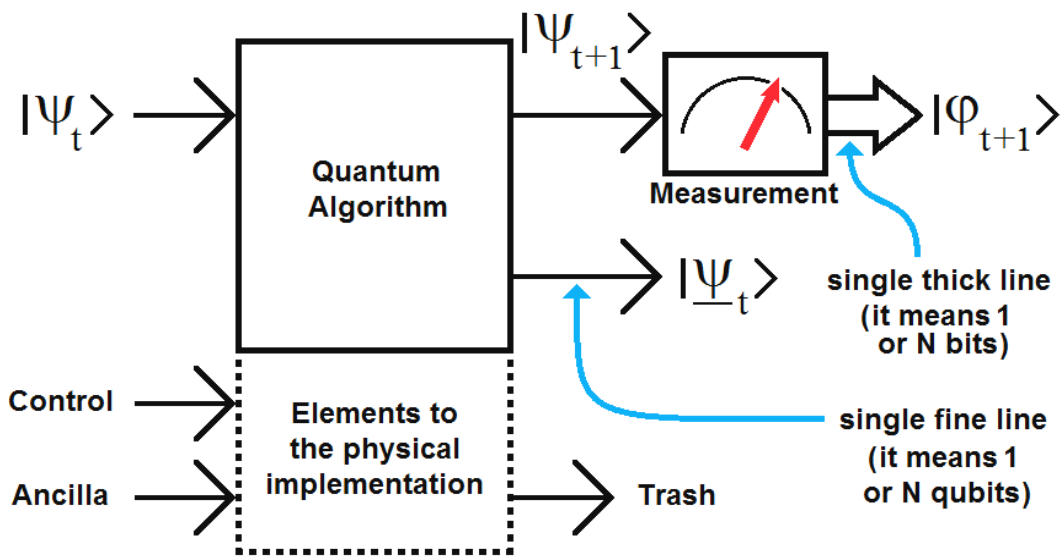

**Fig. 5** Module to measuring, quantum algorithm and the elements needs to its physical implementation.

Besides, we can see clearly a module for measuring $|\psi_{t+1}\rangle$ (which will be extensively discussed in the next section) with their respective output, i.e., $|\varphi_{t+1}\rangle$, and a number of elements needed for the physical imple-mentation of the quantum algorithm (circuit or gate), namely: control, ancilla and trash [1]. In this figure as well as in the rest of them (unlike [1]) a single fine line represents a wire carrying *1* qubit or *N* qubits (qudit), interchangeably, while a single thick line represents a wire carrying *1* or *N* classical bits, interchangeably too.

However, the mentioned concept of reversibility is closely related to energy consumption, and hence to the Landauer's Principle.

On the other hand, computational complexity studies the amount of time and space required to solve a computational problem. Another important computational resource is energy. In this section, we study the energy requirements for computation. Surprisingly, it turns out that computation, both classical and quantum, can in principle be done without expending any energy! Energy consumption in computation turns out to be deeply linked to the reversibility of the computation.

What is the connection between energy consumption and irreversibility in computation? Landauer's principle provides the connection, stating that, in order to erase information, it is necessary to dissipate energy. More precisely, Landauer's principle may be stated as follows:

<u>Landauer's principle (first form):</u> *Suppose a computer erases a single bit of information. The amount of energy dissipated into the environment is at least $k_B T \ln 2$, where $k_B$ is a universal constant known as Boltzmann's constant, and T is the temperature of the environment of the computer.*

According to the laws of thermodynamics, Landauer's principle can be given an alternative form stated not in terms of energy dissipation, but rather in terms of entropy:

<u>Landauer's principle (second form):</u> *Suppose a computer erases a single bit of information. The entropy of the environment increases by at least $k_B \ln 2$, where $k_B$ is Boltzmann's constant.*

Consider a gate like the gate, which takes as input two bits, and produces a single bit as output. This gate is intrinsically irreversible because, given the output of the gate, the input is not uniquely determined. For example, if the output of the gate is 1, then the input could have been any one of 00, 01, or 10. On the other hand, the gate is an example of a reversible logic gate because, given the output of the gate, it is possible to infer what the input must have been. Another way of understanding irreversibility is to think of it in terms of information erasure. If a logic gate is irreversible, then some of the information input to the gate is lost irretrievably when the gate operates – that is, some of the information has been erased by the gate. Conversely, in a reversible computation, no information is ever erased, because the input can always be recovered from the output. Thus, saying that a computation is reversible is equivalent to saying that no information is erased during the computation.

Summing-up, the above expressed justifies the inexcusable need for the presence of $\left| \underline{\psi}_t \right\rangle$ to the output of quantum gate [1].

## 3 Implementation Problems in Quantum Image Processing

In this section, we present the following topics:

- Wave function collapse
- Quantum Measurement Problems
- Before and after measurement
- Types of measurement and state reconstruction
- Interfaces
- Internal representations of an image and its possible implementations

which are fundamental concepts for a posterior development of our own internal representation of an image (inside quantum processor), classical-to-quantum and quantum-to-classical interfaces, and optimal state estimator after quantum measurement.

3.1 Wave function collapse

In quantum mechanics, wave function collapse is the phenomenon in which a wave function -initially in a superposition of several eigenstates- appears to reduce to a single eigenstate after interaction with a measuring apparatus [35]. In Fig. 5, from $|\psi_{t+1}\rangle$ (quantum) to $|\varphi_{t+1}\rangle$ (classical). It is the essence of measurement in quantum mechanics, and connects the wave function with classical observables like position and momentum. Collapse is one of two processes by which quantum systems evolve in time; the other is continuous evolution via the Schrödinger equation [36]. However in this role, collapse is merely a black box for thermodynamically irreversible interaction with a classical environment [37]. Calculations of quantum decoherence predict apparent wave function collapse when a superposition forms between the quantum system's states and the environment's states. Significantly, the combined wave function of the system and environment continue to obey the Schrödinger equation [38].

When the Copenhagen interpretation was first expressed, Niels Bohr postulated wave function collapse to cut the quantum world from the classical [39]. This tactical move allowed quantum theory to develop without distractions from interpretational worries. Nevertheless it was debated, for if collapse were a fundamental physical phenomenon, rather than just the epiphenomenon of some other process, it would mean nature were fundamentally stochastic, i.e. nondeterministic, an undesirable property for a theory [37, 40]. This issue remained until quantum decoherence entered mainstream opinion after its reformulation in the 1980s [37, 38, 41]. Decoherence explains the perception of wave function collapse in terms of interacting large- and small-scale quantum systems, and is commonly taught at the graduate level (e.g. the Cohen-Tannoudji textbook) [42]. The quantum filtering approach [43-46] and the introduction of quantum causality non-demolition principle [47] allows for a classical-environment derivation of wave function collapse from the stochastic Schrödinger equation.

3.2 Quantum Measurement Problems

The measurement problem in quantum mechanics is the unresolved problem of how (or if) wave function collapse occurs. The inability to observe this process directly has given rise to different interpretations of quantum mechanics, and poses a key set of questions that each interpretation must answer. The wave function in quantum mechanics evolves deterministically according to the Schrödinger equation as a linear superposition of different states, but actual measurements always find the physical system in a definite state. Any future evolution is based on the state the system was discovered to be in when the measurement was made, meaning that the measurement "did something" to the process under examination. Whatever that "something" may be does not appear to be explained by the basic theory.

To express matters differently (to paraphrase Steven Weinberg [48, 49]), the Schrödinger wave equation determines the wave function at any later time. If observers and their measuring apparatus are themselves described by a deterministic wave function, why can we not predict precise results for measurements, but only probabilities? As a general question: How can one establish a correspondence between quantum and classical reality? [50].

3.3 Before and after measurement

In quantum mechanics, measurement is a non-trivial and highly counter-intuitive process. Firstly, because measurement outcomes are inherently probabilistic, i.e. regardless of the carefulness in the preparation of a measurement procedure, the possible outcomes of such measurement will be distributed according to a certain probability distribution. Secondly, once a measurement has been performed, a quantum system in unavoidably altered due to the interaction with the measurement apparatus. Consequently, for an arbitrary quantum system, pre-measurement and post-measurement quantum states are different in general [22].

***Postulate.*** Quantum measurements are described by a set of measurement operators $\{\hat{M}_m\}$, index $m$ labels the different measurement outcomes, which act on the state space of the system being measured. Measurement outcomes correspond to values of *observables*, such as position, energy and momentum, which are Hermitian operators [1, 22] corresponding to physically measurable quantities.

Let $|\psi\rangle$ be the state of the quantum system immediately before the measurement. Then, the probability that result $m$ occurs is given by

$$p(m) = \langle \psi | \hat{M}_m^\dagger \hat{M}_m | \psi \rangle \tag{9}$$

and the post-measurement quantum state is

$$|\psi\rangle_{pm} = \frac{\hat{M}_m |\psi\rangle}{\sqrt{\langle \psi | \hat{M}_m^\dagger \hat{M}_m | \psi \rangle}} \tag{10}$$

Operators $\hat{M}_m$ must satisfy the completeness relation, i.e., $\sum_m \hat{M}_m^\dagger \hat{M}_m = I$ [22] because that guarantees that probabilities will sum to one: $\sum_m \langle \psi | \hat{M}_m^\dagger \hat{M}_m | \psi \rangle = \sum_m p(m) = 1$.

Let us work out a simple example. Assume we have a polarized photon with associated polarization orientations 'horizontal' and 'vertical'. The horizontal polarization direction is denoted by $|0\rangle$ and the vertical polarization direction is denoted by $|1\rangle$. Thus, an arbitrary initial state for our photon can be described by the quantum state $|\psi\rangle = \alpha|0\rangle + \beta|1\rangle$, where $\alpha$ and $\beta$ are complex numbers constrained by the normalization condition $|\alpha|^2 + |\beta|^2 = 1$ and $\{|0\rangle, |1\rangle\}$ is the computational basis spanning $H^2$.

Now, we construct two measurement operators $\hat{M}_0 = |0\rangle\langle 0|$ and $\hat{M}_1 = |1\rangle\langle 1|$ and two measurement outcomes $a_0, a_1$. Then, the full observable used for measurement in this experiment is $\hat{M} = a_0 |0\rangle\langle 0| + a_1 |1\rangle\langle 1|$. According to Postulate, the probabilities of obtaining outcome $a_0$ or outcome $a_1$ are given by $p(a_0) = |\alpha|^2$ and $p(a_1) = |\beta|^2$. Corresponding post-measurement quantum states are as follows: if outcome = $a_0$ then $|\psi\rangle_{pm} = |0\rangle$; if outcome = $a_1$ then $|\psi\rangle_{pm} = |1\rangle$.

### 3.4 Types of measurement and state reconstruction

As we have seen in the previous subsection, quantum measurement is not a minor issue [48-50]. In fact, it is an issue still unresolved [51, 52], which would make it impossible for every practical effort to implement any genuine quantum algorithm in general and quantum image processing algorithm in particular. Really, it is an inherited problem of quantum physics and known as the paradox of measurement [53-56].

From a practical point of view, inside context of quantum image processing, the problem is reduced to the following: suppose we develop a quantum algorithm for filtering classic images. A first problem would be (no doubt), how to introduce a classical noisy image within the quantum computer? That is to say, design of the interfaces (classical-to-quantum, and quantum-to classical). But, the second would be, how to measure the results of a quantum filtering algorithm, and to take the result of that filtering process and carry out to the classical world, in other words, the recovery of the classical version of the filtered image into its original space, i.e., the classic world where it was generated. It is obvious that an absolutely accurate technique of measurement is needed. Unfortunately, all efforts in this regard have been useless [57, 58].

However, in the last decade there have been several efforts to remedy this situation, namely:

- Weak measurement
- Restoring the quantum state
- Quantum state tomography

**Weak measurement** is a technique to measure the average value of a quantum *observable* $|\psi\rangle_{pm}$ without appreciably affecting the initial state $|\psi\rangle$ of the system being measured [59-63]. Weak measurements differ from normal (sometimes called "strong" or "von Neumann") measurements in two ways:

1. If $|\psi\rangle_{pm}$ has discrete spectrum (which we assume for simplicity), a strong measurement when the system is in state $|\psi\rangle$ yields an eigenvalue of $|\psi\rangle_{pm}$; if the measurement is repeated many times (starting each time with the system in state $|\psi\rangle$) one obtains a sequence of eigenvalues of $|\psi\rangle_{pm}$ which when averaged yield an approximation to $\langle\psi|\psi_{pm}|\psi\rangle$, the expectation of $|\psi\rangle_{pm}$ in the state $|\psi\rangle$.

By contrast, a *weak measurement* only yields a sequence of numbers which average to $\langle\psi|\psi_{pm}|\psi\rangle$. For example, a strong measurement of the spin of a spin-1/2 particle must yield spin 1/2 or -1/2, but a particular weak measurement could yield spin 100, while a subsequent weak measurement on an identical system might be -128.3 . Typically, a single weak measurement gives little information; only the average of a large number of such measurements is meaningful.

2. A strong measurement changes ("projects") an initial pure state $|\psi\rangle$ to an eigenvector of $|\psi\rangle_{pm}$. (The particular eigenvector obtained cannot be predicted, though its probability is determined.) This substantially changes the state $|\psi\rangle$ unless $|\psi\rangle$ happened to be close to that eigenvector.

However, a weak measurement does not substantially change the initial state.

Weak measurements are usually implemented by coupling the original system $\Psi$ to be measured with an auxiliary quantum "meter system" *M*. The meter along a scale, though in practice various microscopic quantum systems are used. The composite system is mathematically represented as the tensor product of $\Psi$ with *M*, denoted $\Psi \otimes M$. A "product" state in this tensor product is typically denoted $|\psi\rangle|m\rangle$, where $|\psi\rangle$ is a state of $\Psi$ and $|m\rangle$ a state of *M*. States which are not product states are called *entangled* states.

The results obtained by this technique are as weak as its name, therefore, we proceed to the next.

**Restoring the quantum state** is an effort to recover the original state $|\psi\rangle$ from the alleged invertibility of measurement operator through the matrix that represents, that is to say $\hat{M}$ of Section 3.3 [64]. Parrott work is presented in opposition to the technique of weak measurement in general and Katz et al work [65] in particular. Other relevant works mediate between the above [66, 67], also without success.

Today, we know based on Stochastic Processes and Adaptive Filtering [24-31] the single matrix inversion in the process of estimation or identification does not restore the state of a system hidden behind such matrix. This is due to the need to model correctly state and measurement noises and the appropriate architecture of the estimator for the correct system state recovery from the observables. This deficiency explains why Wiener filter was completely replaced by the Kalman's filter in the presence of said noise [24-28]. Therefore, this technique is as weak as that at which it opposes.

**Quantum state tomography** is the process of reconstructing the quantum state (density matrix) for a source of quantum systems by measurements on the systems coming from the source [68, 69]. Being the density matrix for pure or mixed states,

$$\hat{\rho} = \sum_m p(m) |\psi_m\rangle\langle\psi_m| \tag{11}$$

The source may be any device or system which prepares quantum states either consistently into quantum pure states or otherwise into general mixed states. To be able to uniquely identify the state, the measurements must be tomographically complete. That is, the measured operators must form an operator basis on the Hilbert space of the system, providing all the information about the state. Such a set of observations is sometimes called a quorum. In quantum process tomography on the other hand, known quantum states are used to probe a quantum process to find out how the process can be described. Similarly, quantum measurement tomography works to find out what measurement is being performed. The general principle behind quantum state tomography is that by repeatedly performing many different measurements on quantum systems described by identical density matrices, frequency counts can be used to infer probabilities, and these probabilities are combined with Born's rule to determine a density matrix which fits the best with the observations [70, 71]. Obviously, this method is a spartan estimator of the density matrix and not the states themselves. In fact, it is a monitor of the elements of the matrix, only. Therefore, our problem persists.

3.5 Interfaces

Figure 6 shows an overview of compression and decompression of a classic image thanks to the intervention of two quantum processors. The same scheme can be used in the context of filtering and image segmentation.

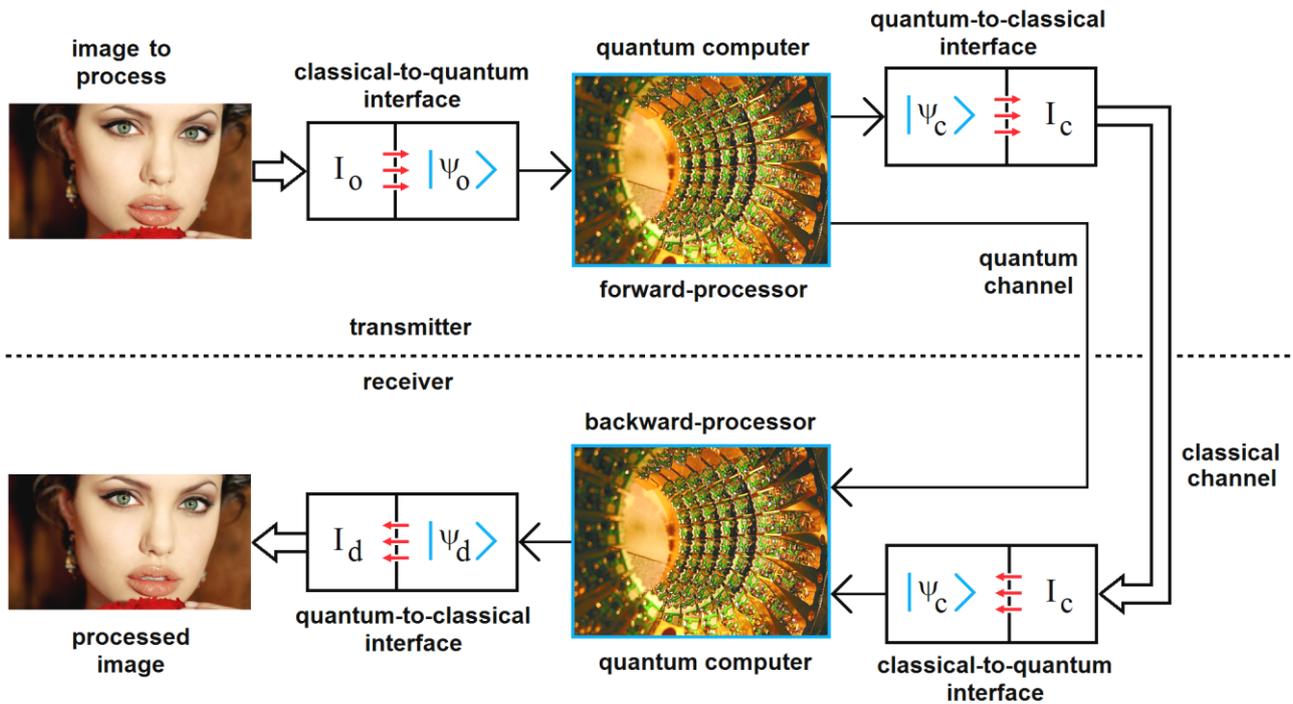

**Fig. 6** General scheme for quantum image processing.

For example, for the case of an image compression context, $I_o$ is the original image and $|\psi_o\rangle$ its counterpart (after classical-to-quantum interface), $|\psi_c\rangle$ is the quantum compression version of the image. If we use a quantum channel, $|\psi_c\rangle$ simply travels from transmitter to receiver. However, if we use a classical channel, it is imperative to use a quantum-to-channel interface, and $I_c$ is the traveling image.

On the other hand, inside receiver, $|\psi_c\rangle$ is decompressed and converted into $|\psi_d\rangle$. Other interface turns $|\psi_d\rangle$ on $I_d$, which is the decompressed classical image.

As we can see, it is absolutely inexcusable to use quantum-to-classical and classical-to-quantum interfaces, which to date is an unresolved problem.

3.6 Internal representations of an image and its possible implementations

In Figure 7, we can see four columns of which the second is the Bloch's sphere. This figure represents an effort to understand how it should be a classic-to-quantum interface, and therefore, the internal representation of the values (for each color, i.e., red-green-blue: RGB) of each pixel of a classic image.

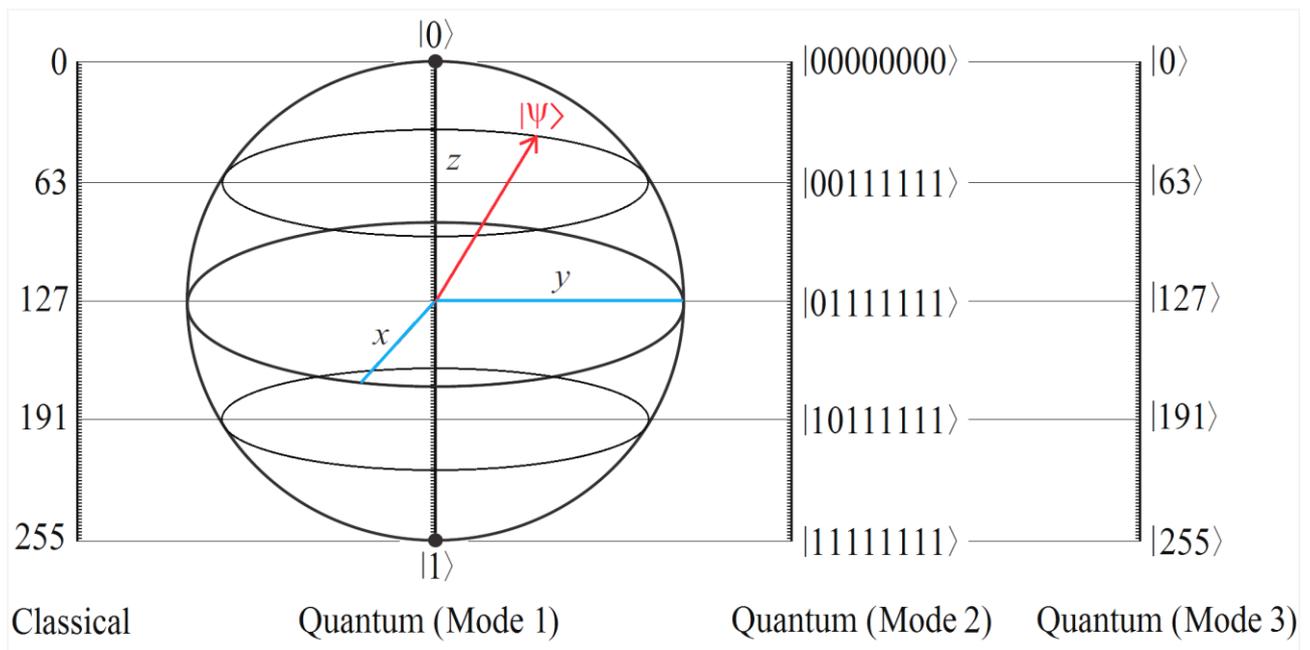

**Fig. 7** Several proposals for internal representations of pixel values for each color.

The first column contains the values of each pixel of the classic image (for each color, i.e., RGB). The second column is the Bloch's sphere with values between $|0\rangle$ and $|1\rangle$. In this column shows three examples of parallel intermediate values associated with its classical counterpart, i.e., 63, 127 and 191. The second column represents the Mode 1 of the internal representation, while the third column represents the Mode 2 and the fourth column represents the Mode 3. Other is to say that none of them provides an efficient internal representation of an image within a quantum computer and much less its processing. For those reasons, another criterion for the internal representation and thus the creation of interfaces and the quantum processing is imperative.

# 4 Pole-to-pole Axis Only (PAO)

According to Equations 1 and 3, as well as, Figures 1 and 2, $\alpha$ is the projection of $|\psi\rangle$ onto axis $z$, i.e.,

$$\alpha = cos\frac{\theta}{2} \tag{12}$$

As we can see in Fig. 8, for $|\psi_r\rangle$ and $|\psi_g\rangle$, $\alpha_r$ and $\alpha_g$ are their projections onto axis $z$, respectively. It is obvious that $|\psi_r\rangle$ has a greater latitude than $|\psi_g\rangle$, i.e., $\alpha_r > \alpha_g$. In other words, $\alpha_r$ is closer $|0\rangle$ (North Pole) than $\alpha_g$. Conversely, $\alpha_r$ is further $|1\rangle$ (South Pole) than $\alpha_g$.

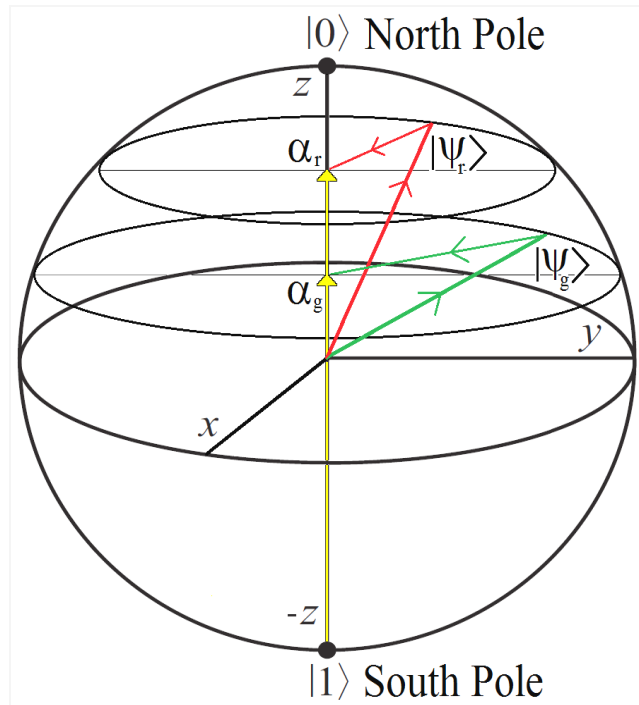

**Fig. 8** Projections onto $z$ axis.

Clearly, the projections on the $x$ and $y$ axis were completely obviated in our previous analysis. This is because such projections are absolutely irrelevant for practical purposes, constituting a principle called Pole-to-pole Axis Only (PAO), whereby, we can represent the pixels of an image (for each color) considering only values from the projections on the z axis.

Based on this simple fact, we can state new criteria, Logic and Arithmetic for the internal representation, processing and interface of images on a quantum computer.

4.1 PAO Criteria

Based on Fig. 9, if we do

$$\mu = 1 - \alpha \tag{13}$$

we can see (for the same example) that $\mu_g > \mu_r$, that is to say, $\mu_r$ is closer $|0\rangle$ (North Pole) than $\mu_g$.

Conversely, $\mu_r$ is further $|1\rangle$ (South Pole) than $\mu_g$. Here seems to work with $\alpha$ and $\mu$ is the same, however, the difference is dramatic. To work with $\mu$ facilitating the internal representation of images, as well as classical-to-quantum and quantum-to-classical interface design and quantum algorithms processing.

Equation 13 receives the name of classical converter.

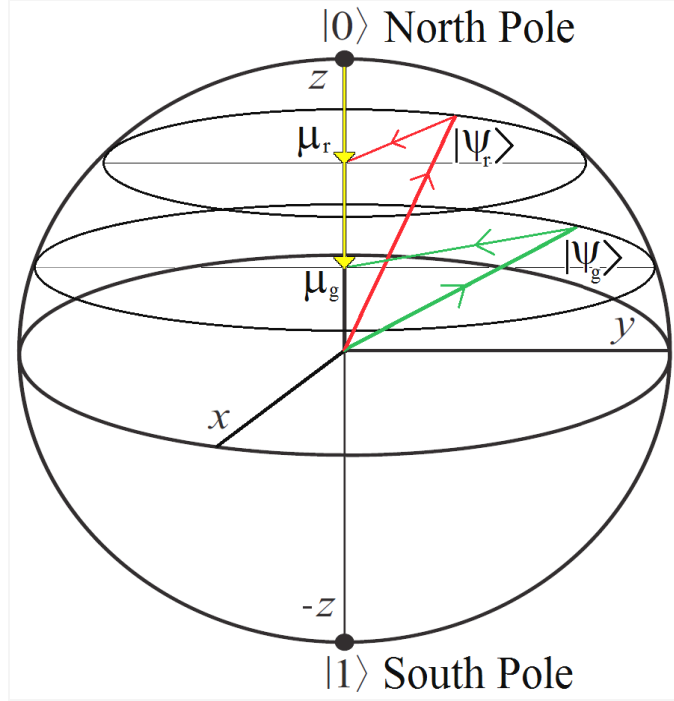

**Fig. 9** Converted projections onto $z$ axis.

Its quantum counterpart is

$$|\mu\rangle = |0\rangle - |\psi\rangle \qquad (14)$$

and it is called quantum converter. Remember that

$$|0\rangle = \begin{bmatrix} 1 \\ 0 \end{bmatrix} \text{ and } |\psi\rangle = \begin{bmatrix} \alpha \\ \beta \end{bmatrix}$$

then

$$\begin{bmatrix} \mu_\alpha \\ \mu_\beta \end{bmatrix} = \begin{bmatrix} 1 \\ 0 \end{bmatrix} - \begin{bmatrix} \alpha \\ \beta \end{bmatrix} = \begin{bmatrix} 1-\alpha \\ -\beta \end{bmatrix} \qquad (15)$$

In this approach the second component is absolutely overlooked, it has the value that it has, i.e., $-\beta$ or any other value, we do not care. Therefore, Equations 13 and 15 coincide, finally. Thus, the POA criteria is:

*It only matters the projection onto z axis, that is, pole-to-pole axis, and their conversions.*

In case of several qubits, for example, in its generic form, we say B qubits, where B = 8 (in most of cases, concerning digital image processing inside quantum processor, and for each color [14-17]). Thus, in the last case, quantum numbers are

$$|00000000\rangle = \begin{pmatrix} 1 \\ 0 \\ \vdots \\ 0 \end{pmatrix} \Big\} 256\, elements \qquad (16)$$

$$|11111111\rangle = \begin{pmatrix} 0 \\ 0 \\ \vdots \\ 1 \end{pmatrix} \Big\} 256\, elements \qquad (17)$$

In both Equations, (16) and (17), each column has 1 *one* and 255 *zeros*. Otherwise, these numbers can be represented as $|0\rangle$ and $|2^B - 1\rangle = |255\rangle$ (with B = 8), respectively. In such case, $\alpha$ limits will be $min(\alpha) = 0$ and $max(\alpha) = 2^B - 1$, i.e., $0 \leq \alpha \leq 2^B - 1$. Therefore, classic converter is

$$\mu = (2^B - 1) - \alpha \qquad (18)$$

while the quantum converter is

$$|\mu\rangle = |0\rangle - |\psi\rangle \qquad (19)$$

That is to say, the same of Equation (14), however, $|0\rangle$ is associated with 256 levels instead of 2.

4.2 Logic operations according to PAO Criteria

For computational basis states ($|0\rangle, |1\rangle$), we can set four basic logic operations (AND, OR, XOR, and NOT), where $\overline{(\cdot)}$ means $NOT(\cdot)$, see Table I.

TABLE I
LOGIC OPERATIONS FOR COMPUTATIONAL BASIS STATES

| $|\psi_1\rangle$ | $|\psi_2\rangle$ | AND | OR | $\overline{|\psi_1\rangle}$ | $\overline{|\psi_2\rangle}$ | XOR |
|---|---|---|---|---|---|---|
| $|0\rangle$ | $|0\rangle$ | $|0\rangle$ | $|0\rangle$ | $|1\rangle$ | $|1\rangle$ | $|0\rangle$ |
| $|0\rangle$ | $|1\rangle$ | $|0\rangle$ | $|1\rangle$ | $|1\rangle$ | $|0\rangle$ | $|1\rangle$ |
| $|1\rangle$ | $|0\rangle$ | $|0\rangle$ | $|1\rangle$ | $|0\rangle$ | $|1\rangle$ | $|1\rangle$ |
| $|1\rangle$ | $|1\rangle$ | $|1\rangle$ | $|1\rangle$ | $|0\rangle$ | $|0\rangle$ | $|0\rangle$ |

being $|\psi_i\rangle = \begin{pmatrix} \alpha_i \\ \beta_i \end{pmatrix}$

Therefore,

$$\overline{|\psi_i\rangle} = NOT(|\psi_i\rangle) = NOT\begin{pmatrix} \alpha_i \\ \beta_i \end{pmatrix} = \begin{pmatrix} \beta_i \\ \alpha_i \end{pmatrix} \quad \forall i \qquad (20)$$

with

$$XOR(|\psi_1\rangle, |\psi_2\rangle) = (|\psi_1\rangle \text{ AND } \overline{|\psi_2\rangle}) \text{ OR } (\overline{|\psi_1\rangle} \text{ AND } |\psi_2\rangle) \tag{21}$$

In traditional Quantum Logic. these four operations (AND, OR, XOR, and NOT) are implemented using the Control-NOT (CNOT) and Toffoli gates [1]. On the other hand, we can see what happens with $\alpha$ and $\mu$ in these cases.. In Table II, we have,

TABLE II
$\alpha$ AND $\mu$ FOR COMPUTATIONAL BASIS STATES

| $\alpha_1$ | $\alpha_2$ | $\mu_1$ | $\mu_2$ | AND | OR | XOR |
|---|---|---|---|---|---|---|
| 1 | 1 | 0 | 0 | $|0\rangle$ | $|0\rangle$ | $|0\rangle$ |
| 1 | 0 | 0 | 1 | $|0\rangle$ | $|1\rangle$ | $|1\rangle$ |
| 0 | 1 | 1 | 0 | $|0\rangle$ | $|1\rangle$ | $|1\rangle$ |
| 0 | 0 | 1 | 1 | $|1\rangle$ | $|1\rangle$ | $|0\rangle$ |

However, regarding to $\mu$, logics operations seems Boolean operations, see Table III. In fact, as we can see in Fig. 10, when result of AND operation is $|0\rangle$, one of the two $\mu$ equals to 0 and we are on North Pole, while in the case where AND operation is $|1\rangle$, for both $\mu$ equal to 1 and we are on South Pole.

TABLE III
LOGIC OPERATIONS REGARDING $\mu$

| $\mu_1$ | $\mu_2$ | $AND_\mu$ | $OR_\mu$ | $XOR_\mu$ |
|---|---|---|---|---|
| 0 | 0 | 0 | 0 | 0 |
| 0 | 1 | 0 | 1 | 1 |
| 1 | 0 | 0 | 1 | 1 |
| 1 | 1 | 1 | 1 | 0 |

Conversely, when result of OR operation is $|0\rangle$, for both $\mu$ equal to 0 and we are on North Pole, while in the case where OR operation is $|1\rangle$, with only one of them is 1. That is to say, inside PAO Criteria, AND operation is a minimum, while OR operation is a maximum between both $|\psi\rangle$. This is extended beyond the pure case of computational basis states ($|0\rangle, |1\rangle$). Then, we obtain

$$\begin{aligned} |\psi_1\rangle \wedge |\psi_2\rangle &= min(|\psi_1\rangle, |\psi_2\rangle) \\ |\psi_1\rangle \vee |\psi_2\rangle &= max(|\psi_1\rangle, |\psi_2\rangle) \\ |\psi_1\rangle \veebar |\psi_2\rangle &= max\left(min(|\psi_1\rangle, \overline{|\psi_2\rangle}), min(\overline{|\psi_1\rangle}, |\psi_2\rangle)\right) \end{aligned} \tag{22}$$

where, $\wedge$ means AND, $\vee$ means OR, and $\veebar$ means XOR.

Here we can draw some conclusions:

- these logical operations for qubits placed serve anywhere in the Bloch's sphere (even if not pure qubits), thing that traditional quantum logic gates can not do [1, 72, 73].

- consistent with PAO criteria, it only imports projections on the vertical axis (pole-to-pole), i.e., it only care to know which is the highest or lowest parallel, in other words, the northernmost and southernmost parallel. Summing-up, we will consider only the projections on the vertical axis in the measurement process.

- the AND and OR logic operations give the same results as those obtained by the same operations in Fuzzy Logic [74-80]. i.e., *min*(.) and *max*(.), respectively. The allows us to imagine future applications of this technology applying it to Automatic Control [81, 82].

- then, this new logic is only possible after an exact measurement (without change on state, or wave collapse). If such thing is possible (thank to the estimator to be presented in this work) we can compare and order quantum states, which is impossible today [83, 84].

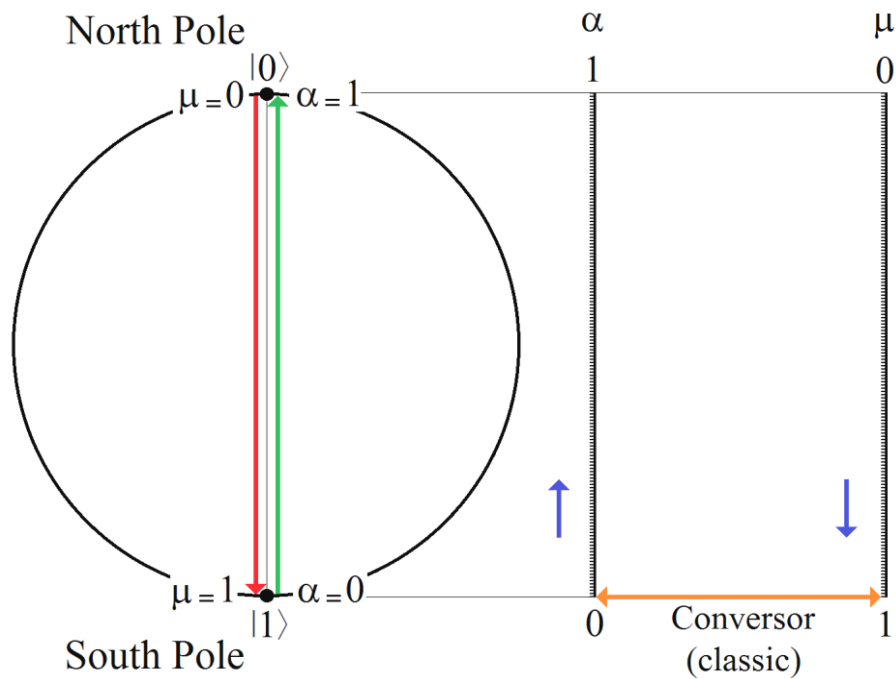

**Fig. 10** $\alpha$, $\mu$ and logic operation results on Bloch's sphere.

This new logic is better even than Multivalued Quantum Logic [85].

4.3 Arithmetic operations according to PAO Criteria

Consistent with this and from here it requires simple arithmetic quantum operations.

*For 1 qubit:*

being $|\psi_1\rangle = \begin{pmatrix} \alpha_1 \\ \beta_1 \end{pmatrix}$ and $|\psi_2\rangle = \begin{pmatrix} \alpha_2 \\ \beta_2 \end{pmatrix}$, the *Traditional Quantum Arithmetic* [86-94] is composed of

- Quantum Addition [1, 95-110]:

$$|\psi_1\rangle + |\psi_2\rangle = \begin{pmatrix} \alpha_1 \\ \beta_1 \end{pmatrix} + \begin{pmatrix} \alpha_2 \\ \beta_2 \end{pmatrix} = \begin{pmatrix} \alpha_1 + \alpha_1 \\ \beta_1 + \beta_2 \end{pmatrix} \qquad (23)$$

- Quantum Substraction [1]:

$$|\psi_1\rangle - |\psi_2\rangle = \begin{pmatrix} \alpha_1 \\ \beta_1 \end{pmatrix} - \begin{pmatrix} \alpha_2 \\ \beta_2 \end{pmatrix} = \begin{pmatrix} \alpha_1 - \alpha_1 \\ \beta_1 - \beta_2 \end{pmatrix} \qquad (24)$$

- Quantum Multiplication [111-114]:

$$|\psi_1\rangle \times |\psi_2\rangle = \begin{pmatrix} \alpha_1 \\ \beta_1 \end{pmatrix} \times \begin{pmatrix} \alpha_2 \\ \beta_2 \end{pmatrix} = \begin{pmatrix} \alpha_1 \alpha_2 \\ \beta_1 \alpha_2 \\ \alpha_1 \beta_2 \\ \beta_1 \beta_2 \end{pmatrix} \qquad (25)$$

- Quantum Division [115, 116]:

$$|\psi_1\rangle / |\psi_2\rangle = \begin{pmatrix} \alpha_1 \\ \beta_1 \end{pmatrix} / \begin{pmatrix} \alpha_2 \\ \beta_2 \end{pmatrix} = \begin{pmatrix} \alpha_1 / \alpha_2 \\ \beta_1 / \alpha_2 \\ \alpha_1 / \beta_2 \\ \beta_1 / \beta_2 \end{pmatrix} \qquad (26)$$

So the *New Quantum Arithmetic* according to Equation (19) for $|\mu_1\rangle = |0\rangle - |\psi_1\rangle = \begin{pmatrix} 1 \\ 0 \end{pmatrix} - \begin{pmatrix} \alpha_1 \\ \beta_1 \end{pmatrix} = \begin{pmatrix} 1 - \alpha_1 \\ -\beta_1 \end{pmatrix}$, and $|\mu_2\rangle = |0\rangle - |\psi_2\rangle = \begin{pmatrix} 1 \\ 0 \end{pmatrix} - \begin{pmatrix} \alpha_2 \\ \beta_2 \end{pmatrix} = \begin{pmatrix} 1 - \alpha_2 \\ -\beta_2 \end{pmatrix}$, it will be to apply the same equations developed above but with these changes of variables, i.e., $|\mu_1\rangle$ instead of $|\psi_1\rangle$, $|\mu_2\rangle$ instead of $|\psi_2\rangle$,

- Addition:

$$|\mu_1\rangle + |\mu_2\rangle = \begin{pmatrix} 1 - \alpha_1 \\ -\beta_1 \end{pmatrix} + \begin{pmatrix} 1 - \alpha_2 \\ -\beta_2 \end{pmatrix} = \begin{pmatrix} 2 - (\alpha_1 + \alpha_2) \\ -(\beta_1 + \beta_2) \end{pmatrix} \qquad (27)$$

- Substraction:

$$|\mu_1\rangle - |\mu_2\rangle = \begin{pmatrix} 1 - \alpha_1 \\ -\beta_1 \end{pmatrix} - \begin{pmatrix} 1 - \alpha_2 \\ -\beta_2 \end{pmatrix} = \begin{pmatrix} \alpha_2 - \alpha_1 \\ \beta_2 - \beta_1 \end{pmatrix} \qquad (28)$$

- Multiplication:

$$|\mu_1\rangle \times |\mu_2\rangle = \begin{pmatrix} 1-\alpha_1 \\ -\beta_1 \end{pmatrix} \times \begin{pmatrix} 1-\alpha_2 \\ -\beta_2 \end{pmatrix} = \begin{pmatrix} (1-\alpha_1)(1-\alpha_2) \\ (-\beta_1)(1-\alpha_2) \\ (1-\alpha_1)(-\beta_2) \\ (-\beta_1)(-\beta_2) \end{pmatrix} \quad (29)$$

- Division:

$$|\mu_1\rangle / |\mu_2\rangle = \begin{pmatrix} 1-\alpha_1 \\ -\beta_1 \end{pmatrix} / \begin{pmatrix} 1-\alpha_2 \\ -\beta_2 \end{pmatrix} = \begin{pmatrix} (1-\alpha_1)/(1-\alpha_2) \\ (-\beta_1)/(1-\alpha_2) \\ (1-\alpha_1)/(-\beta_2) \\ (-\beta_1)/(-\beta_2) \end{pmatrix} \quad (30)$$

In 4 cases it only interest us the first component of the result vector (PAO criteria).

*For B qubits (qudit)*

The results are similar in shape to the previous case. Two things changed:

- we replace 1 by $2^B-1$
- the size of the qubits involved is substantially higher. However, here also, it only interest us the projection onto the pole-to-pole axis of the result vector (PAO criteria).

Fig. 11 show us the relationship between $\alpha$ and $\mu$ and the final processed $I$ (for each color), specifically, two examples, the red one ($\alpha_r$, $\mu_r$), and the green one ($\alpha_g$, $\mu_g$).

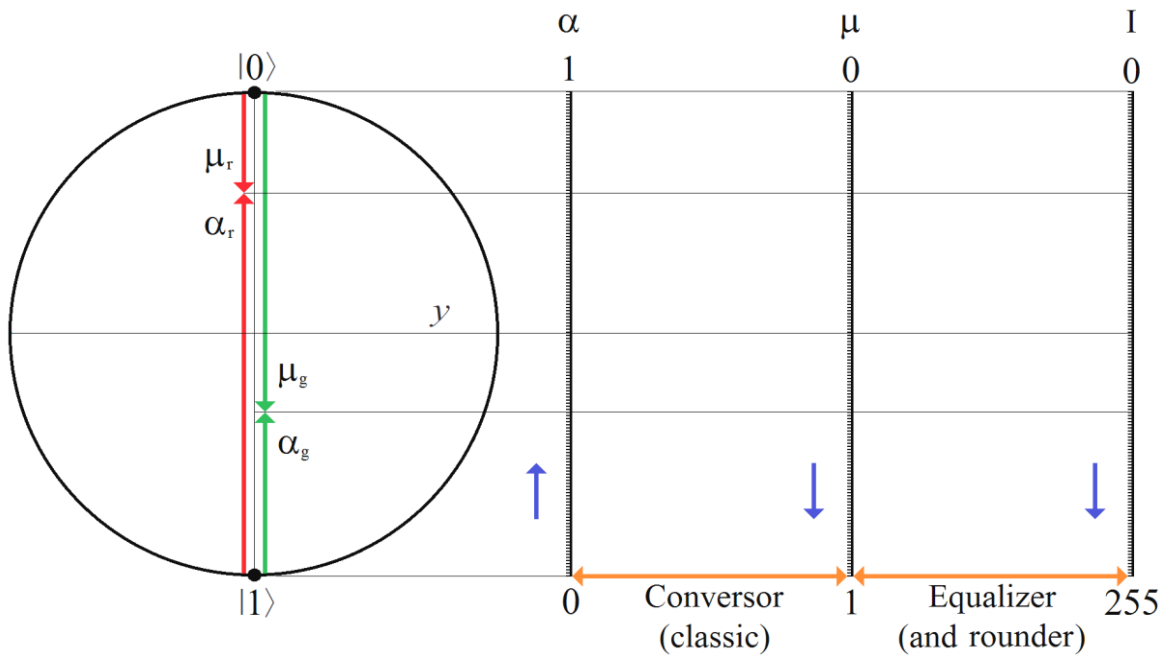

**Fig. 11** Relationship between $\alpha$ and $\mu$ and the final processed $I$ (via equalizer and rounder).

Three important details to note are:

- the progress of *I* coincides with the progress of $\mu$, fully backward to $\alpha$.
- the scales of $\mu$ and hence *I* are nonlinear
- it is necessary an equalizer and rounder between $\mu$ and hence *I*

## 5 Optimal State Estimator (OSE)

### 5.1 Classical state estimator in noiseless environments

In order to develop an optimal estimate of quantum states, we start defining everything on a classical type of estimator called Recursive Least Squre RLS [29-31] and derived from the famous Kalman's filter [24-28]. Such estimator (time discrete version and in noiseless environment) is based on Fig. 12, in which,

A: plant $\in \mathbb{R}^{N \times N}$
M: measurement operator $\in \mathbb{R}^{M \times N}$
$\Delta$: unitary delay $(N \times N)$
t: time
X: state to be estimated $\in \mathbb{R}^{N \times 1}$
Y: observable $\in \mathbb{R}^{M \times 1}$
$\varepsilon$: error of estimation $\in \mathbb{R}^{M \times 1}$
K: Kalman's gain $\in \mathbb{R}^{N \times M}$
$\hat{X}$ : estimated state $\in \mathbb{R}^{N \times 1}$
$\hat{Y}$ : output of estimator $\in \mathbb{R}^{M \times 1}$

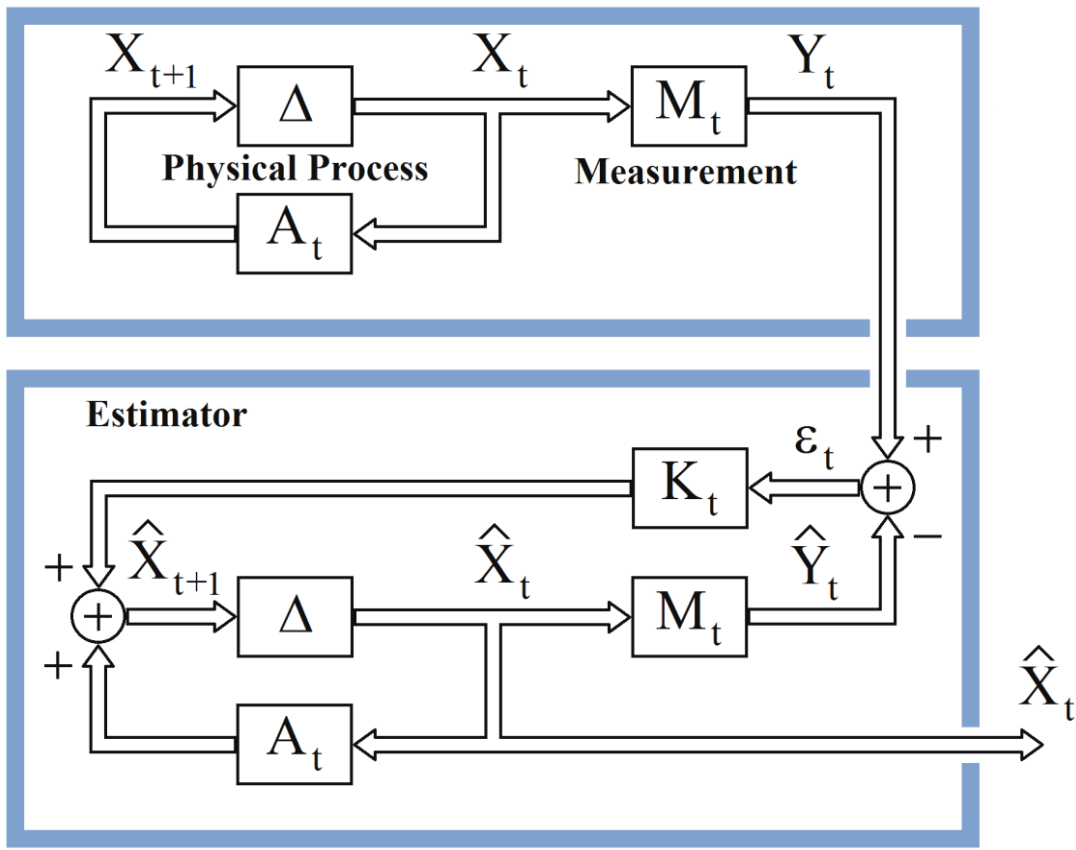

**Fig. 12** RLS.

*Original System:*

$$X_t = A_t X_{t-1} \tag{31}$$
$$Y_t = M_t X_t \tag{32}$$

*Estimator:*

$$\hat{X}_t = A_t \hat{X}_{t-1} + K_t \varepsilon_t \tag{33}$$
$$\hat{Y}_t = M_t \hat{X}_t \tag{34}$$

We can then define *a priori* and *a posteriori* (respectively) estimate error as:

$$\varepsilon_t^- = Y_t - \hat{Y}_t^- = Y_t - M_t \hat{X}_t^- \tag{35}$$

and

$$\varepsilon_t = Y_t - \hat{Y}_t = Y_t - M_t \hat{X}_t \tag{36}$$

The *a priori* estimate error covariance is then

$$\Xi\left\{(\varepsilon_t^-)(\varepsilon_t^-)^T\right\} = \Xi\left\{(Y_t - M_t \hat{X}_t^-)(Y_t - M_t \hat{X}_t^-)^T\right\} \tag{37}$$

where $\Xi\{\bullet\}$ means square error of "$\bullet$", and $(\bullet)^T$ means transpose of "$(\bullet)$".

On the other hand, the *a posteriori* estimate error covariance is

$$\Xi\{\varepsilon_t \varepsilon_t^T\} = \Xi\left\{(Y_t - M_t \hat{X}_t)(Y_t - M_t \hat{X}_t)^T\right\} \tag{38}$$

This adaptation process is based on the minimization of the mean square error criterion defined in the last equation. Developing Equation (38), rearranging terms, and minimizing the mean square error with respect to $\hat{X}$, we obtain the Wiener filter to stationary signals, that is to say,

$$\hat{X} = R_{MM}^{-1} r_{MY} \tag{39}$$

where, $R_{MM}$ is the autocorrelation matrix M and $r_{MY}$ is the cross-correlation vector of M and Y. In the following, we formulate a recursive, time-update, adaptive formulation of Equation (39). In fact, $R_{MM}$ can be expressed in recursive fashion as

$$R_{MM,t} = R_{MM,t-1} + M_t M_t^T \tag{40}$$

To introduce adaptability to the time variations of the signal statistics, the autocorrelation estimate in Equation (40) can be windowed by an exponentially decaying window:

$$R_{MM,t} = \lambda R_{MM,t-1} + M_t M_t^T \tag{41}$$

where $\lambda$ is the so-called adaptation, or forgetting factor, and is in the range $0 < \lambda < 1$. Similarly, the cross-correlation vector can be calculated in recursive form as

$$r_{MY,t} = r_{MY,t-1} + M_t Y_t \tag{42}$$

Again this equation can be made adaptive using an exponentially decaying forgetting factor $\lambda$:

$$r_{MY,t} = \lambda r_{MY,t-1} + M_t Y_t \tag{43}$$

For a recursive solution of the least square error Equation (43), we need to obtain a recursive time-update formula for the inverse matrix in the form

$$R_{MM,t}^{-1} = R_{MM,t-1}^{-1} + \text{Update}_t \tag{44}$$

where "Update$_t$" is an update factor to be actualized in each step time. After an extensive series of considerations, developments and replacements (such as $P_{MM,t} = R_{MM,t}^{-1}$), we get the follo-wing set of equations related to RLS adaptation algorithm [29-31] (very similar to Kalman's filter [24-28]).

Initial values:

- $P_{MM,0} = \delta I$ (being I the identity matrix and $\delta$ a number different to 0)

- $\hat{X}_0 = \hat{X}_I$

Filter gain matrix:

$$K_t = P_{MM,t-1}^{-} M_t \left[ \lambda I + M_t^T P_{MM,t-1}^{-} M_t \right]^{-1} \tag{45}$$

Error signal equation:

$$\varepsilon_t^{-} = Y_t - M_t \hat{X}_{t-1}^{-} \tag{46}$$

Estimated states

$$\hat{X}_t = \hat{X}_{t-1}^{-} - K_t \varepsilon_t^{-} \tag{47}$$

Inverse correlation matrix update:

$$P_{MM,t} = \lambda^{-1} \left[ I - K_t M_t \right] P_{MM,t-1}^{-} \tag{48}$$

Discrete estimator time update equations

$$\hat{X}_t^{-} = A_t \hat{X}_{t-1} \tag{49}$$

$$P_{MM,t-1}^{-} = A_t P_{MM,t-1} A_t^T \tag{50}$$

Indeed, A and M are time-invariant [24-31]. In fact, we can dispense with the Equation (50).

5.2 Quantum state estimator in noiseless environments

From Equation (10), we have

$$|\psi\rangle_{pm} = |\varphi\rangle = \frac{\hat{M}_m |\psi\rangle}{\sqrt{\langle \psi | \hat{M}_m^\dagger \hat{M}_m |\psi\rangle}} \qquad (51)$$

being $\sqrt{\langle \psi | \hat{M}_m^\dagger \hat{M}_m |\psi\rangle}$ a norm of $\hat{M}_m$, as follow,

$$\|\hat{M}_m\| = \sqrt{\langle \psi | \hat{M}_m^\dagger \hat{M}_m |\psi\rangle} \qquad (52)$$

In fact, we can take any norm of $\hat{M}_m$, even for different $|\psi\rangle$ of the original. Thus,

$$|\varphi\rangle = \frac{\hat{M}_m}{\|\hat{M}_m\|} |\psi\rangle = \tilde{M}_m |\psi\rangle \qquad (53)$$

for each $m$, i.e., a battery of estimators, as show in Fig. 13.

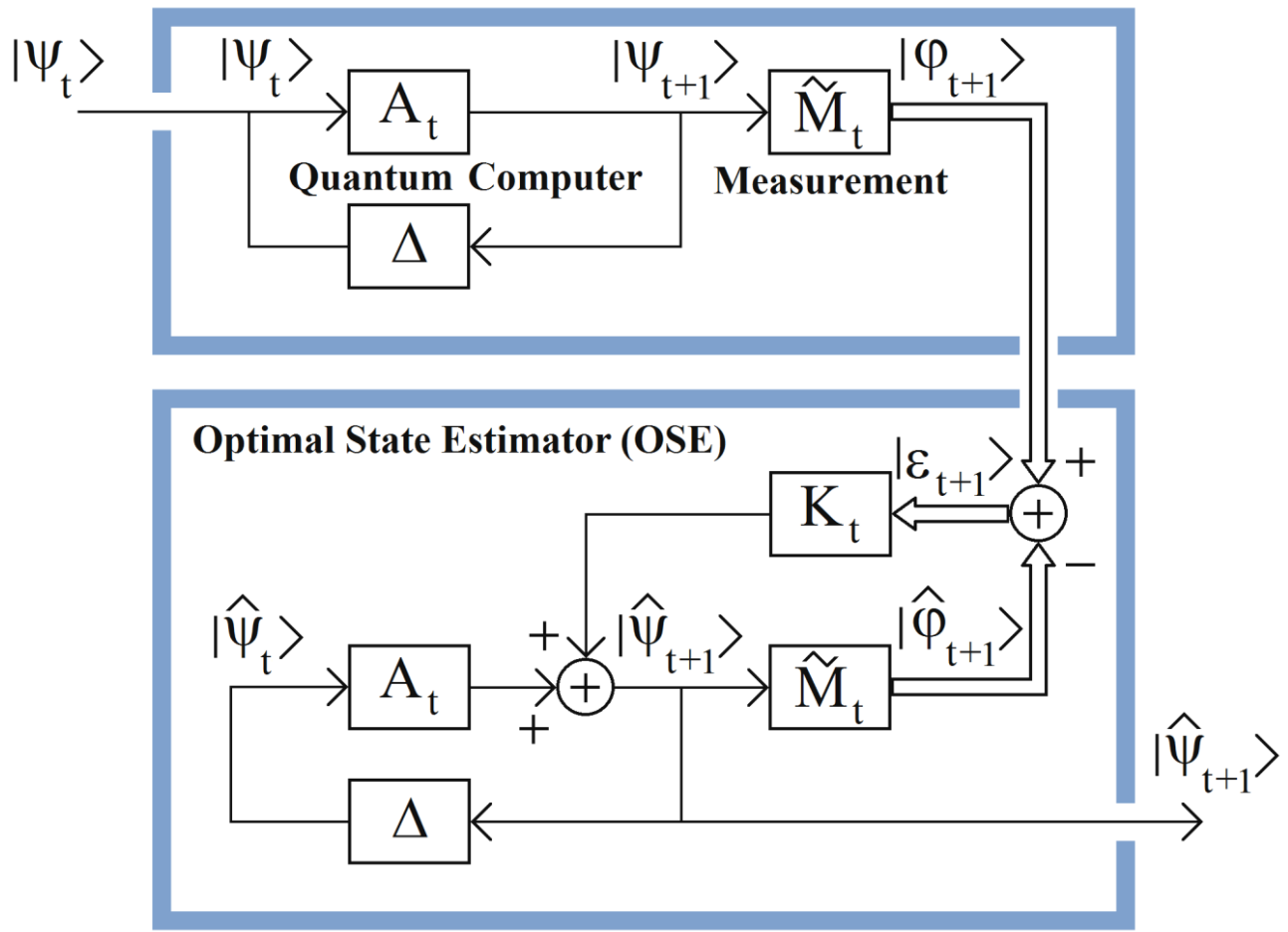

**Fig. 13** Modified RLS.

According to PAO, we measure only the projections on the $z$ axis, i.e., the $\alpha$'s, being A the quantum algorithm (circuit or gate). In fact, we can get $\alpha$ and $\beta$ for each $m$ with this estimator, however, we are only interested $\alpha$'s, therefore, the estimator becomes the same as the classic case, being able to naturally apply PAO. Besides, by PAO we can avoid the Kronecker's Algebra [1] inside estimator.

Based on Fig. 13, the complete set of equations is,

*Inside Quantum Computer:*

$$|\psi_{t+1}\rangle = A_t |\psi_t\rangle \quad \text{(quantum algorithm)} \tag{54}$$

$$|\varphi_{t+1}\rangle = \tilde{M}_t |\psi_{t+1}\rangle \quad \text{(quantum measurement)} \tag{55}$$

*Optimal State Estimator (OSE):*

$$|\hat{\psi}_{t+1}\rangle = A_t |\hat{\psi}_t\rangle + K_t |\varepsilon_{t+1}\rangle \tag{56}$$

$$|\hat{\varphi}_{t+1}\rangle = \tilde{M}_t |\hat{\psi}_{t+1}\rangle \tag{57}$$

*Estimation error:*

$$|\varepsilon_{t+1}\rangle = |\varphi_{t+1}\rangle - |\hat{\varphi}_{t+1}\rangle \tag{58}$$

Three important considerations:

- indeed, A is time-invariant,
- really, OSE is a reorganized RLS/Kalman's filter, but it's the same algorithmically,
- we started with a poor measurement and evolution of OSE improves the accuracy of measurement

As PAO, we can work in two modes, namely:

Mode I:

$$|\underline{\psi}_t\rangle = |0\rangle - |\psi_t\rangle \tag{59}$$

Based on Fig. 13, the estimated and measured $\alpha$ is the final value, which is linked to the final image through a process of equalization and rounding, see next section.

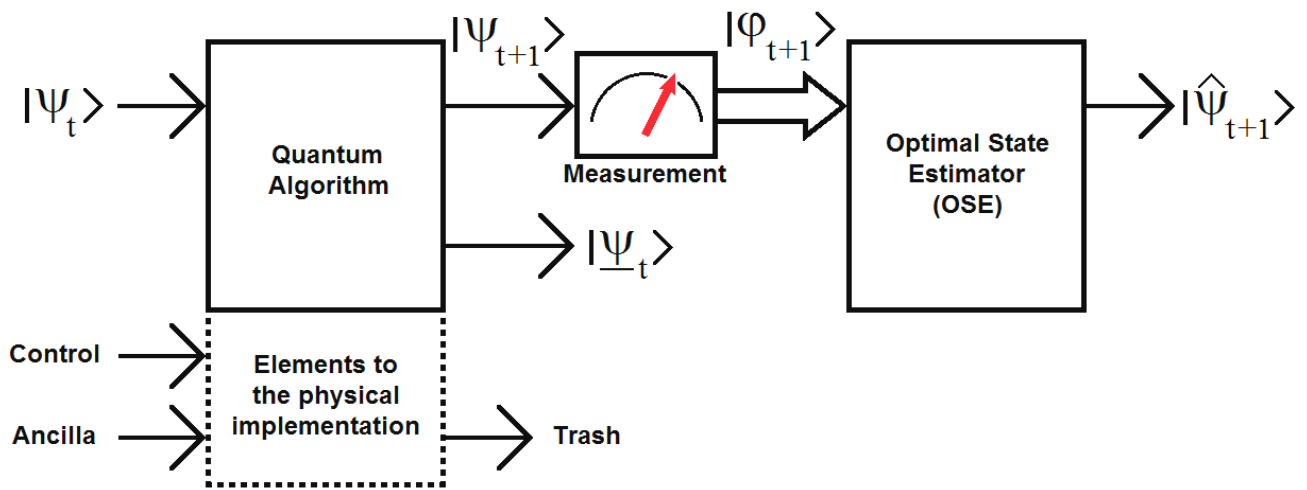

**Fig. 14** Quantum algorithm (circuit or gate), measurement and OSE.

Mode II:

$$|\psi_t\rangle = |\psi_t\rangle \tag{60}$$

with Equation (13) as main reference, that is to say, to the estimated and measured value of alpha $\alpha$, we will apply Equation (13). As in the previous mode, it only remains to equalize and to round $\mu$.

Figure 14 shows the complete schematic of Figure 5 but now with the OSE added to its output.

5.3 Quantum state estimator in noisy environments

We assume the existence of state and measurement noise, as seen in Fig. 15, with equation inside quantum computer

$$|\psi_{t+1}\rangle = A_t |\psi_t\rangle + N^s_{t+1} \quad \text{(quantum algorithm)} \tag{61}$$

$$|\varphi_{t+1}\rangle = \tilde{M}_t |\psi_{t+1}\rangle + N^m_{t+1} \quad \text{(quantum measurement)} \tag{62}$$

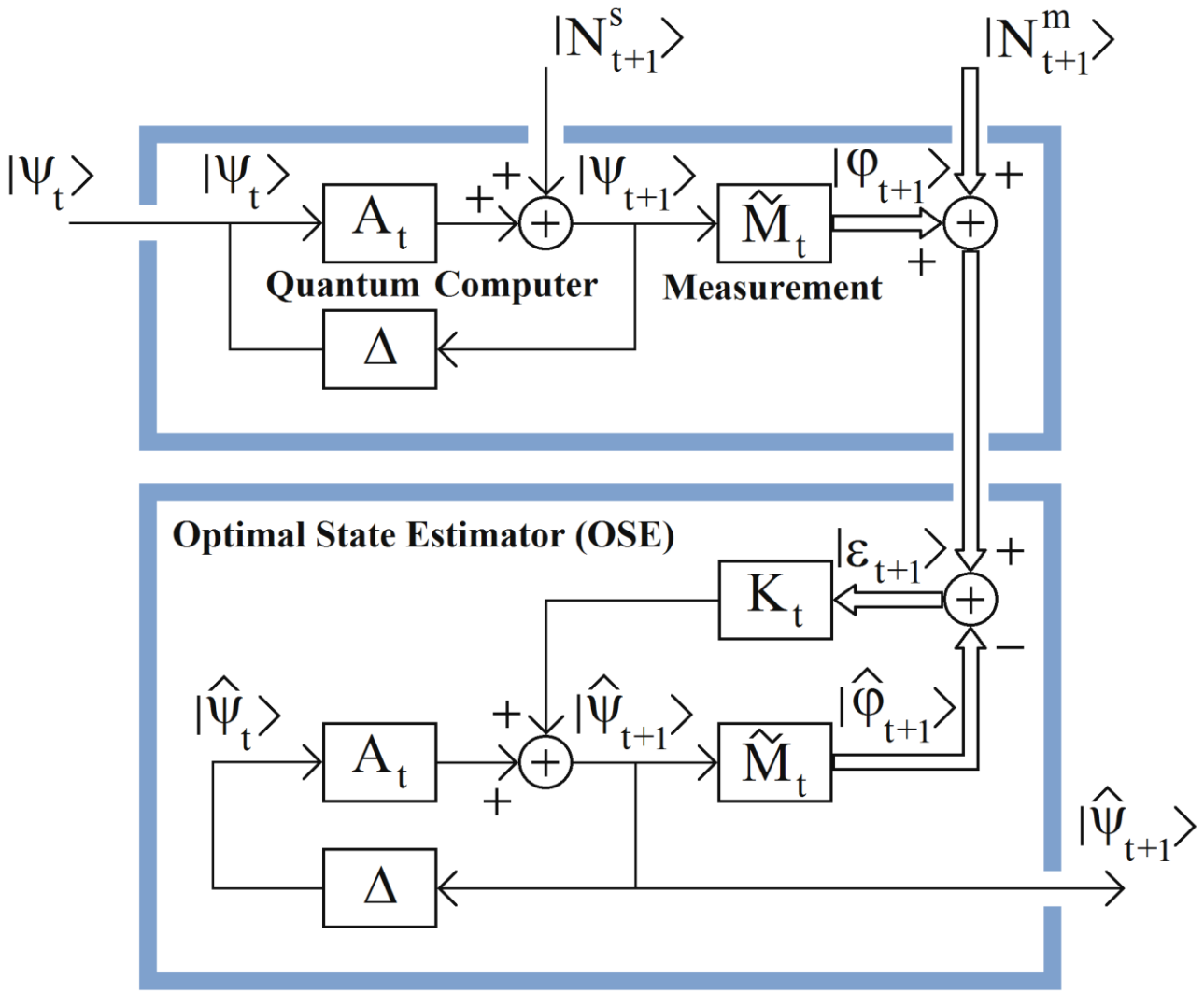

**Fig. 15** Modified Kalman's estimator for noisy environments.

where, the random variables $N_{t+1}^s$ and $N_{t+1}^m$ represent the state and measurement noise, respectively. Both are assumed to be independent (of each other). In practice, the state noise covariance Q, and measurement noise covariance R matrices might change with each time step or measurement, however here we assume both are constant. Thus, only three equations change regarding to classic estimator, namely,

Filter gain matrix:

$$K_t = P_{MM,t-1}^- M_t \left[ R + M_t^T P_{MM,t-1}^- M_t \right]^{-1} \tag{63}$$

Inverse correlation matrix update:

$$P_{MM,t} = \left[ I - K_t M_t \right] P_{MM,t-1}^- \tag{64}$$

Discrete estimator time update equation

$$P_{MM,t-1}^- = A_t P_{MM,t-1} A_t^T + Q \tag{65}$$

However, and as the OSE is a linear system, we can move state noise to the output and work with a unique noise that represents both. Therefore, the last equation is not used.

All these noises may be associated with different factors: quantum noise [1, 22, 87, 94, 117], quantum decoherence [22, 118-123], and measurement errors [48-71]. The accuracy of our estimator (OSE) depends on two aspects:

- our ability to model these noises
- the greater or lesser presence of such noise in the experiment

## 6 Proposed Interfaces

Building interfaces between the classical and quantum world is an aspiration within this field [135-141]. Major obstacle in the construction of such interfaces are:

- decoherence [22, 118-123]
- functional interpretation of such an interface should work. This is the greatest contribution of PAO.

*What is quantum decoherence?*

In quantum mechanics, quantum decoherence is the loss of coherence or ordering of the phase angles between the components of a system in a quantum superposition. One consequence of this dephasing is classical or probabilistically additive behavior. Quantum decoherence gives the appearance of wave function collapse (the reduction of the physical possibilities into a single possibility as seen by an observer) and justifies the framework and intuition of classical physics as an acceptable approximation: decoherence is the mechanism by which the classical limit emerges from a quantum starting point and it determines the location of the quantum-classical boundary. Decoherence occurs when a system interacts with its environment in a thermodynamically irreversible way. This prevents different elements in the quantum superposition of the total scene's wavefunction from interfering with each other. Decoherence has been a subject of active research since the 1980s.

Decoherence can be viewed as the loss of information from a system into the environment (often modeled as a heat bath), since every system is loosely coupled with the energetic state of its surroundings. Viewed in isolation, the system's dynamics are non-unitary (although the combined system plus environment evolves in a unitary fashion). Thus the dynamics of the system alone are irreversible. As with any coupling, entanglements are generated between the system and environment. These have the effect of sharing quantum information with—or transferring it to—the surroundings.

Decoherence does not generate actual wave function collapse. It only provides an explanation for the observation of wave function collapse, as the quantum nature of the system "leaks" into the environment. That is, components of the wavefunction are decoupled from a coherent system, and acquire phases from their immediate surroundings. A total superposition of the global or universal wavefunction still exists (and remains coherent at the global level), but its ultimate fate remains as an interpretational issue. Specifically, decoherence does not attempt to explain the measurement problem. Rather, decoherence provides an explanation for the transition of the system to a mixture of states that seem to correspond to those states observers perceive. Moreover, our observation tells us that this mixture looks like a proper quantum ensemble in a measurement situation, as we observe that measurements lead to the "realization" of precisely one state in the "ensemble".

Decoherence represents a challenge for the practical realization of quantum computers, since such machines are expected to rely heavily on the undisturbed evolution of quantum coherences. Simply put, they require to preserve coherent states and to manage decoherence, in order to actually perform quantum computation. In short, the quantum decoherence is a trivial decoherence and unwanted passage between the quantum and the classical world.

*What is functional interpretation of such an interface should work?*

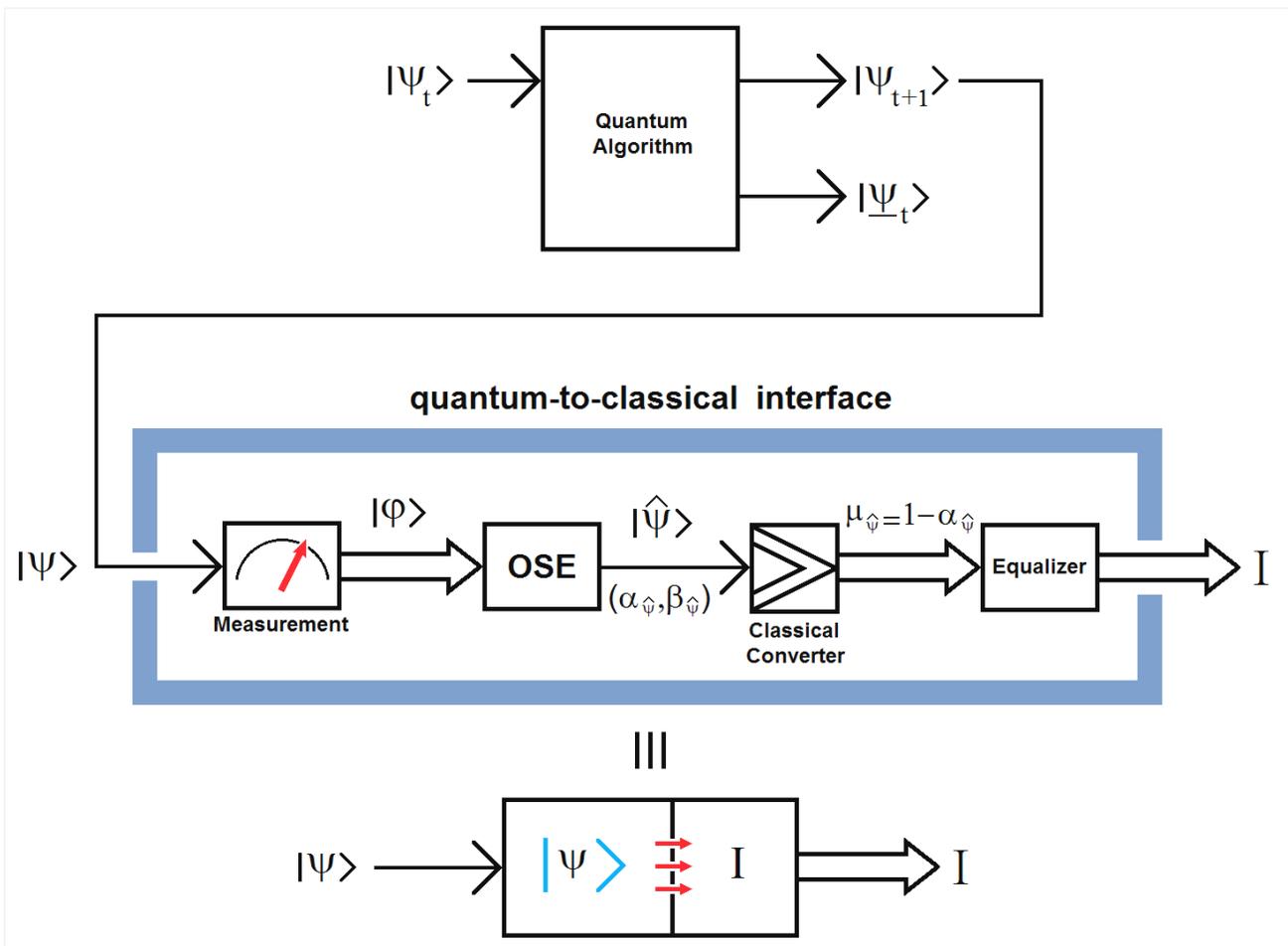

**Fig. 16** Quantum-to-classical interface (Mode I).

The answer to this question lies in the possibility offered PAO as a functional instrument that conveys building these interfaces. In this section, we developed two modes of quantum-to-classical interface and one mode of classical-to-quantum interface according to the above.

6.1 Quantum-to-classical interface (Mode I)

In Fig. 16, we can see first the quantum algorithm, whose output is directed to the interface, which begins with the measurement operator, continuous with the OSE, and thanks to the first mode of PAO recovered $\mu_{\hat{\psi}}$, from $\alpha_{\hat{\psi}}$, so we use a classical converter, with $\mu_{\hat{\psi}} = (2^B - 1) - \alpha_{\hat{\psi}}$ for B qubits. Finally, we employed an equalizer and a rounder (which is not in the figure) to complete the scheme. This architecture is extensible to any dimension of qudits.

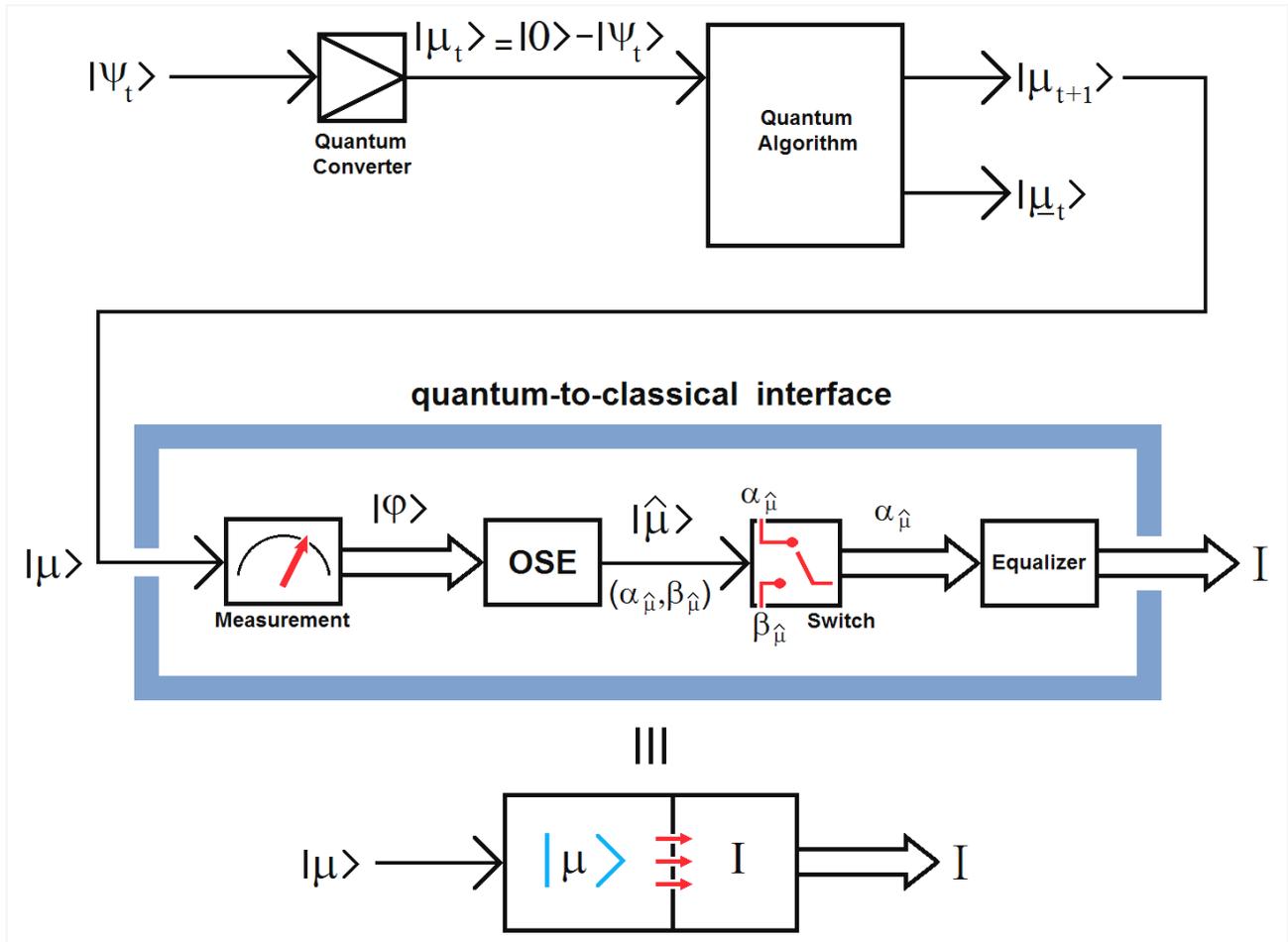

Fig. 17 Quantum-to-classical interface (Mode II).

6.2 Quantum-to-classical interface (Mode II)

In Fig. 17, we can see first a quantum converter, then the quantum algorithm, whose output is directed to the interface, which begins with the measurement operator, continuous with the OSE, and thanks to the second mode of PAO recovered directly $\mu$, of which we are only interested in the projection on the vertical axis, i.e., $\alpha$, so we use the switch. Finally, we employed an equalizer and a rounder to complete the scheme. This architecture is extensible to any dimension of qudits. The $|0\rangle$ of quantum converter is for $2^B = 256$ levels, i.e., B = 8 qubits. In other words, everything said here is for qubits and qudits interchangeably.

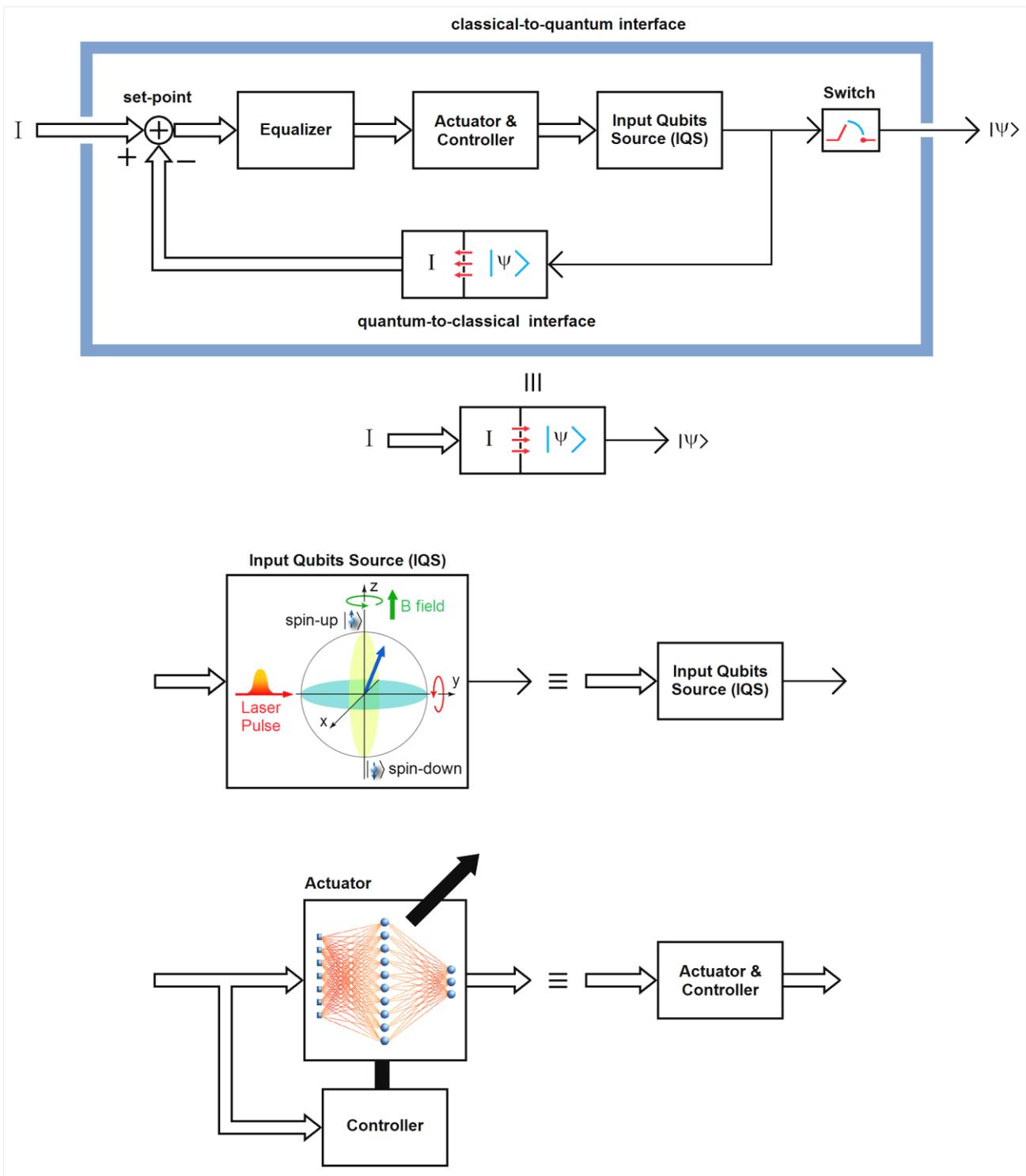

**Fig. 18** Classical-to-quantum interface.

6.3 Classical-to-quantum interface

As we can see in Fig. 18, this interface is essentially an automatic control system, with a set-point (for feedback) whose output goes to the equalizer. Hence the signal flow goes to the actuator and controller, and finally, to the Input Qubit Source (IQS), that is to say, the qubits factory. To the IQS output, we make a measurement by a quantum-to-classical interface. When the level of vertical component of qubits matches the corresponding value of the image (i.e., steady-state is reached) then the switch will close, and the qubits will pass to the input of an eventual quantum algorithm (circuit or gate).

# 7 A bit of Digital Image Processing

In this section, we present a pair of methods -exclusively- from Digital Image Processing [14-17], they are denoising/despeckling (i.e., noise or speckle filtering, respectively) and edge detection. We present here their classical versions alone. Considering the above features of quantum arithmetic and PAO criteria, the extension to the quantum version of these two methods is automatic. However, some preliminary considerations are necessary.

## 7.1 Preliminaries

Despeckling is the process designed to remove the speckle [124-132]. But, what is speckle? Speckle is the noise appearing in Synthetic Aperture Radar (SAR) images [133, 134]. Speckle is usually modelled as a purely multiplicative noise process of the form

$$\begin{aligned} I_s(r,c) &= I(r,c).S(r,c) \\ &= I(r,c).[1+T(r,c)] \\ &= I(r,c)+N(r,c) \end{aligned} \quad (66)$$

The true radiometric values of the image are represented by *I*, and the values measured by the radar instrument are represented by $I_s$. The speckle noise is represented by *S*. The parameters *r* and *c* means row and column of the respective pixel of the image. If *S(r,c) = 1+T(r,c)* and *N(r,c) = I(r,c).T(r,c)*, one begins with a multiplicative speckle *S* and finishes with an additive speckle *N* [124-132], which avoids the log-transform, because the mean of log-transformed speckle noise does not equal to zero [133] and thus requires correction to avoid extra distortion in the restored image.

For single-look SAR images, *S* is Rayleigh distributed (for amplitude images) or negative exponentially distributed (for intensity images) with a mean of *1*. For multi-look SAR images with independent looks, *S* has a gamma distribution with a mean of *1*. Further details on this noise model are given in [134].

Hence, despeckling is considered as a critical preprocessing step in medical imaging systems, SAR imagery among others.

Speckle noise follows a gamma distribution and is given as in following Fig. 19

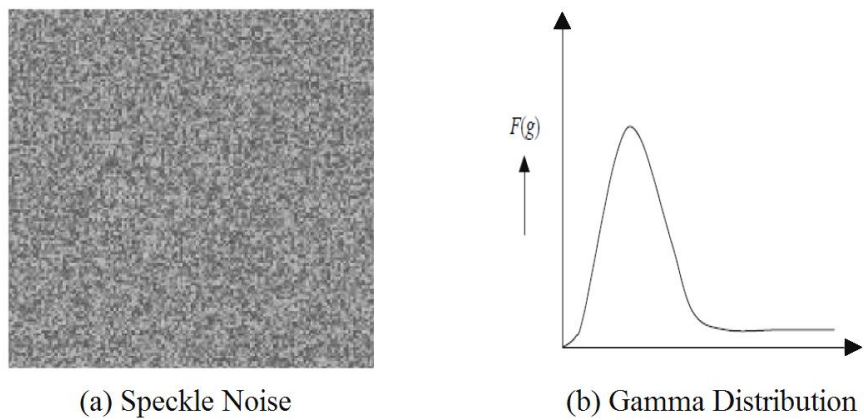

(a) Speckle Noise        (b) Gamma Distribution

**FIG. 19** Speckle and its distribution.

$$F(g) = \frac{g^{\alpha-1} e^{-\frac{g}{a}}}{(\alpha-1)! a^{\alpha}} \tag{67}$$

where variance is $a^{\alpha}$ and $g$ is the gray level. On an image, speckle noise (with variance 0.05) looks as shown in Fig. 19a and the corresponding gamma distribution is given in Fig. 19b.

7.2 Convolutive mask

In both cases (i.e., filtering and edge detection) we use algorithms based on a convolutive masks with a horizontal rafter (see Fig. 20) on the noisy image [124-126], and on that image to which we must make an edge detection.

In the filtering case, we employ a mask for directional smoothing [124-126], whereas in edge detection we use the Sobel procedure [14-17].

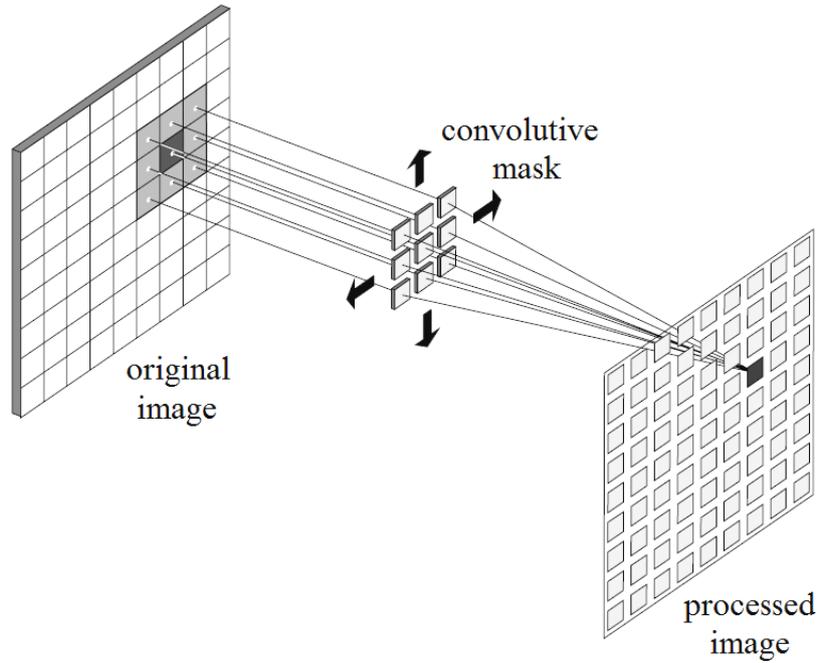

**Fig. 20** The convolution between the mask and the original image in a horizontal rafter produce the processed image.

*Directional Smoothing (DS):*

To protect the edges from blurring while smoothing, a directional averaging filter must be applied. Spatial averages $d(r,c:\Theta)$ are calculated in several directions as shown in the following equation

$$d(r,c:\Theta) = \frac{1}{N_{\Theta}} \sum_{k \in W_{\Theta}} \sum_{l \in W_{\Theta}} I(r-k, c-l) \tag{68}$$

and a direction $\Theta^*$ is found such that $|I(r,c) - d(r,c:\Theta^*)|$ is minimum, where $I$ is the respective image. Then

$$d(r,c) = d(r,c : \Theta^*) \qquad (69)$$

gives the desired result for the suitably chosen window $W$, $N_\Theta$ is the number of directions, and $k$ and $l$ depends on the size of such windows (kernel) [14-17].

The DS filter has a speckle reduction approach that performs spatial filtering in a square-moving window defined as kernel, and is based on the statistical relationship between the central pixel and its surrounding pixels as shown in Fig. 21.

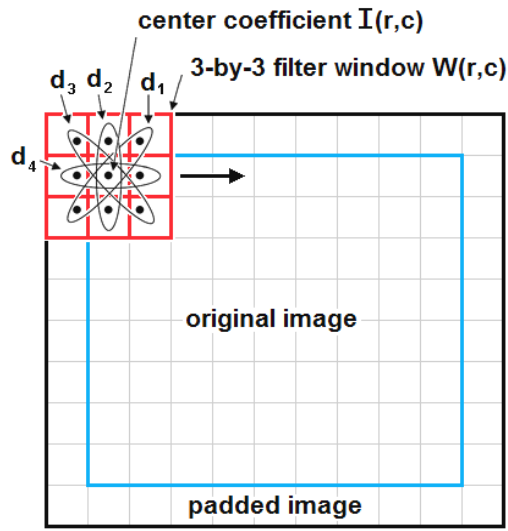

**Fig. 21** 3-by-3 filter window for DS on a Image.

The size of the filter window can range from 3-by-3 to 33-by-33, with an odd number of cells in both directions. A larger filter window means that a larger area of the image will be used for calculation and requires more computation time depending on the complexity of the filter's algorithm. If the size of the filter window is too large, the important details will be lost due to over smoothing. On the other hand, if the size of the filter window is too small, speckle reduction may not be very effective. In practice, a 3-by-3 or a 7-by-7 filter window usually yields good results in the cases under study [14-17].

DS performs the filtering based on either local statistical data given in the filter window to determine the noise variance within the filter window, or estimating the local noise variance.

*Sobel Procedure:*

Sobel filtering is a *three step* process. Two 3×3 filters (often called *kernels*) are applied separately and independently, see Fig. 22.

The weights these kernels apply to pixels in the 3×3 region are depicted below:

$$k_H = \begin{bmatrix} -1 & 0 & +1 \\ -2 & 0 & +2 \\ -1 & 0 & +1 \end{bmatrix}, \text{ horizontal kernel} \qquad (70)$$

and

$$k_v = \begin{bmatrix} +1 & +2 & +1 \\ 0 & 0 & 0 \\ -1 & -2 & -1 \end{bmatrix}, \text{vertical kernel} \tag{71}$$

Again, notice that in both cases, the sum of the weights is 0. The idea behind these two filters is to approximate the derivatives in x and y, respectively. Call the results of these two filters $D_x(x, y)$ and $D_y(x, y)$. Both $D_x$ and $D_y$ can have positive or negative values, so you need to add 0.5 so that a value of 0 corresponds to a middle gray in order to avoid clamping (to [0..1]) of these intermediate results.

The final step in the Sobel filter approximates the gradient magnitude based on the partial derivatives ($D_x(x, y)$ and $D_y(x, y)$) from the previous steps. The gradient magnitude, which is the result of the Sobel Filter $S(x, y)$, is simply:

$$S(x,y) = \sqrt{\left(D_x(x,y)\right)^2 + \left(D_y(x,y)\right)^2} \tag{72}$$

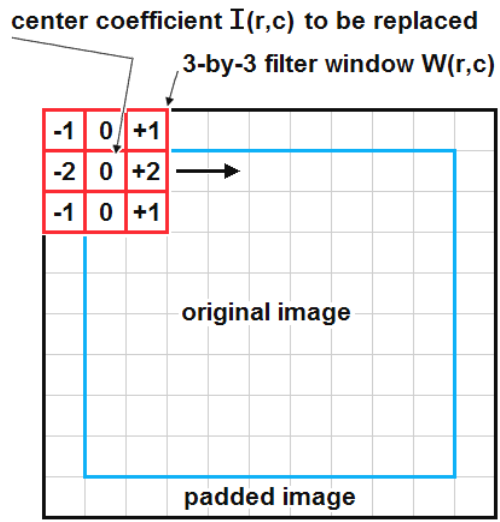

**Fig. 22** 3x3 filter window for Sobel edge-detection on a Image.

Please note that your textures should not store $D_x(x, y)$, but should rather store $D_y(x, y)$ +0.5, as I mentioned above. This means before computing the value in Equation (72), you need to first subtract the 0.5 you added when computing $D_x$ and $D_y$.

So, in summary, the three steps are:

- Compute the image storing partial derivatives in x ($D_x(x, y)$) by applying the right 3×3 kernel to the *original* input image.

- Compute the image storing partial derivatives in y ($D_y(x, y)$) by applying the left 3×3 kernel to the *original* input image.

- Compute the gradient magnitude $S(x, y)$ based on $D_x$ and $D_y$.

Two further things to notice about Sobel filters: (*a*) both the derivative kernels depicted above are separable, so they could be split into disjoint x and y passes, and (*b*) the entire filter can actually be implemented in a single-pass filter in a relatively straightforward manner.

Here, the filters to be used in the simulations in both versions, classical and quantum.

# 8 Metrics and Simulations

In this section, we present a set of metrics especially designed for experiments which are developed here, and which consists in the comparison of classical and quantum version of denoising/despeckling and edge detection algorithms, outside and inside quantum computer, respectively.

## 8.1 Metrics

Metrics presented here were specially designed for this work.

*Mean Absolute Difference (MAD)*
This is a conspicuous metric for these cases, which it is a quantity used to measure how close forecasts or predictions are to eventual outcomes. The mean absolute difference (MAD) for gray scale images is given by

$$MAD = \frac{\sum_{r,c} |I_{classical}(r,c) - I_{quantum}(r,c)|}{R \times C} \qquad (73)$$

which for two $R \times C$ (rows-by-columns) images $I_{classical}$ and $I_{quantum}$, $I_{classical}$ means classical processed image, and $I_{quantum}$ means quantum processed image.

*Mean Square Difference (MSD)*
*MSD* indicates average square difference of the pixels throughout the image between the classical processed image $I_{classical}$ and the quantum processed image $I_{quantum}$, see Figures 15 and 16. A lower *MSD* indicates a smaller difference between both images. This means that there is a significant filter concordance. Nevertheless, it is necessary to be very careful with the edges. The formula for the *MSD* calculation for gray scale images is

$$MSD = \frac{\sum_{r,c} (I_{classical}(r,c) - I_{quantum}(r,c))^2}{R \times C} \qquad (74)$$

Here $R \times C$ pixels is the size of the images too, including original image *I*.

*Peak Classical-To-Quantum Ratio (PCQR)*
PCQR is a term for the ratio between the maximum possible power of an $I_{classical}$ and the power of corrupting difference that affects the fidelity of the quantum representation regarding classical representation. Because many $I_{classical}$ have a very wide dynamic range, PCQR will be expressed in terms of the logarithmic decibel scale.

We will use it as a measure of quality of coincidence between classical and quantum versions. It is most easily defined via the mean squared difference (MSD) which for two $R \times C$ (rows-by-columns) gray scale images $I_{classical}$ and $I_{quantum}$, that is to say:

$$PCQR = 10 \log_{10} \left( \frac{max(I_{classical})^2}{MSD} \right) = 20 \log_{10} \left( \frac{max(I_{classical})}{\sqrt{MSD}} \right) \qquad (75)$$

Here, $max(I_{classical})$ is the maximum pixel value of the image. When the pixels are represented using 8 bits per sample, this is 255. More generally, when samples are represented using linear pulse code modulation

(PCM) with $B$ bits per sample, maximum possible value of $max(I_{classical})$ is $2^B-1$. For color images with three red-green-blue (RGB) values per pixel, the definition of PCQR is the same except the MSD is the sum over all squared value differences divided by image size and by three.

Typical values for the PCQR are between 30 and 50 dB, where higher is better.

8.2 Simulations

Simulations are organized in two big groups:

- Denoising (and despecklig), see Fig. 23, where $I_n$ means classical noisy (or speckled) image, $|\psi_n\rangle$ is the quantum noisy (or speckled) image, $|\psi_d\rangle$ is quantum denoised (and despeckled) image, $I_{d,q}$ is the classical denoised (and despeckled) image from quantum filtering process, and $I_{d,c}$ is the classical denoised (and despeckled) image from a classical filtering process.

- Edge detection (Fig. 24) where I is the original classical image, $|\psi\rangle$ is the quantum image, $|\psi_s\rangle$ is the quantum edge detected image, $I_{s,q}$ is the classical edge detected image from quantum edge-detection process, and $I_{s,c}$ is the classical edge detected image from a classical edge-detection process.

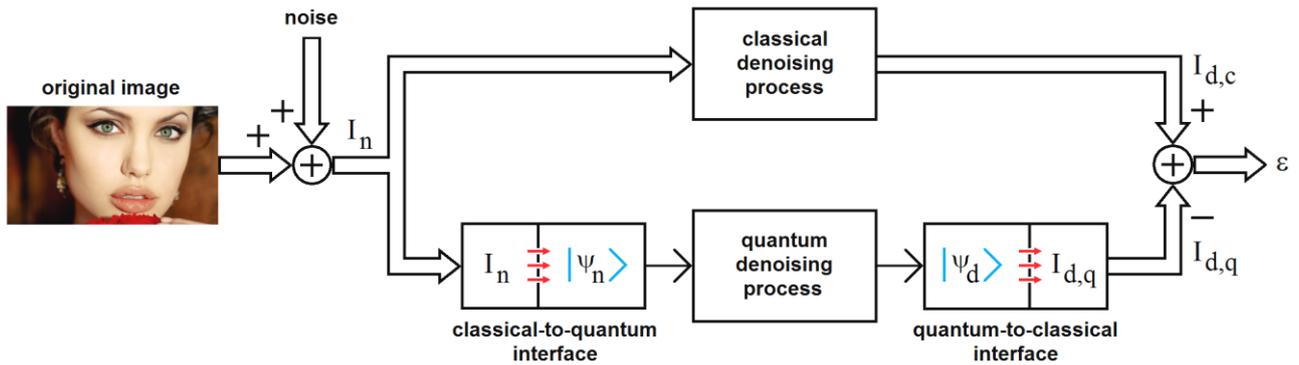

**Fig. 23** Denoising/despckling for classical and quantum contexts.

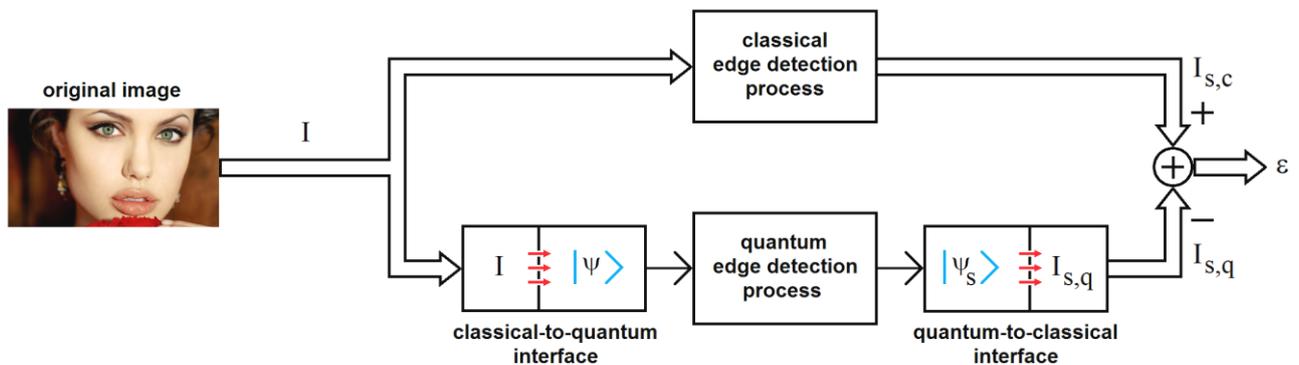

**Fig. 24** Edge detection for classical and quantum contexts.

*Despeckling of a SAR image*

Here, a set of experimental results using one ERS SAR Precision Image (PRI) standard of Buenos Aires city is presented. Fig. 25 (top-left) shows a speckled image used in this experiment from remote sensing satellite ERS-2, with a 1524-by-1524 (pixels) by 256 (gray levels); and the filtered images, processed by using classical directional smoothing (down-left), and quantum directional smoothing techniques (down-right), respectively. Besides, Fig. 25 (top-right) shows the difference pixel-to-pixel between classical and quantum versions. As we can see, there are values of pixels where the difference between two versions is remarkably sensitive.

On the other hand, the original image has no compression, that is to say, the PRI SAR image comes directly from the lossless raw-data SAR processor, in this particular case, Chirp-Scaling [133, 134].

Table IV shows the results of metrics (MAD, MSD and PCQR) for this experiment. Although PQCR is greater than 30 db, however, there is a noticeable difference between the classical and quantum versions of directional smoothing.

TABLE IV
METRICS OF DESPECKLING IN SAR IMAGERY: CLASSICAL VS. QUANTUM

| MAD | MSD | PCQR |
|---|---|---|
| 6.4205 | 64.0690 | 30.0643 |

If the algorithms are essentially the same, the only reason for this difference is some type of noise not correctly modeled in the OSE, either state or measurement.

*Denoising of multimedia images*

First image is *Agus in Miami* (Fig. 26), which is a color Bitmap File Format (lossless) [142] of 1326-by-1326 pixels with 24 bit-per-pixel (bpp). Noise was generated using a MATLAB® R2014a (Mathworks, Natick, MA) [143] built-in function called *imnoise*. The noise type was *salt & pepper*, with a noise density of 0.05.

Fig. 26 (top-left) shows us the original image used in this experiment; noisy image (top-right); the filtered images, processed by using classical directional smoothing (middle-left), and quantum directional smoothing techniques (middle-right), respectively. Besides, Fig. 26 (down-center) shows the difference pixel-to-pixel between classical and quantum versions. As we can see, there are values of pixels where the difference between two versions is remarkably sensitive here too.

In Table V, we can see that PCQR is better than SAR case, however, it is a low value. This is telling us a mismatch between the classical representation of the directional smoothing and its quantum counterpart, which can only be due to improper modeling of noise involved.

TABLE V
METRICS OF DENOISING: CLASSICAL VS. QUANTUM

| IMAGE | MAD | MSD | PCQR |
|---|---|---|---|
| AGUS | 2.7331 | 23.1847 | 34.4788 |
| ANGELINA | 2.4259 | 19.4260 | 35.2470 |
| LENA | 2.4279 | 19.2057 | 35.2965 |

From Digital Image Processing [14-17], we know that such a mismatch (in this type of metric, for example, Mean Square Error) only indicates serious problems with the representation or process modeling. In the two

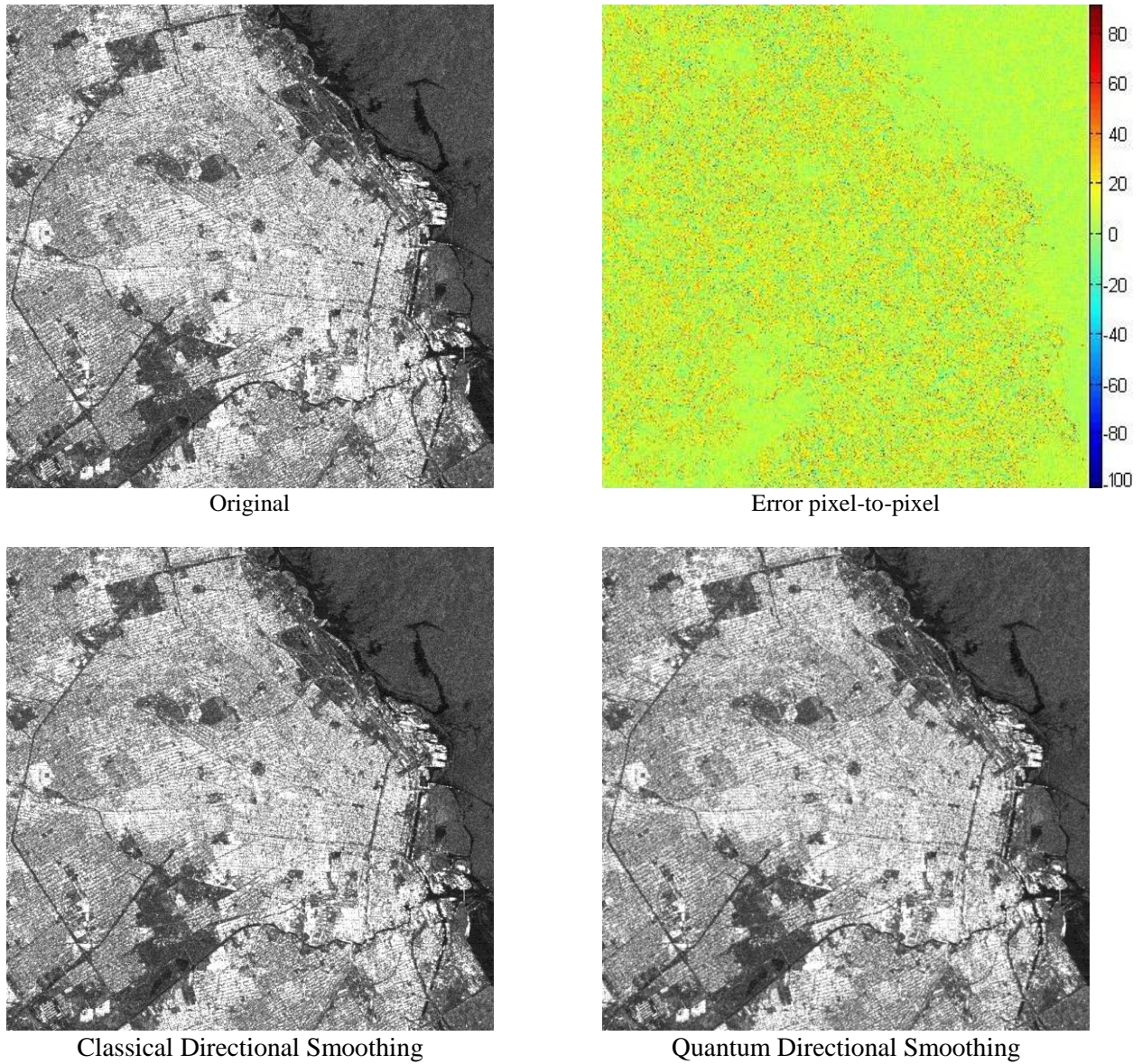

Original        Error pixel-to-pixel

Classical Directional Smoothing        Quantum Directional Smoothing

**Fig. 25** Despeckling for SAR ERS-2 of Buenos Aires City.

previous examples it can be seen in the high MSD values, that is to say, 64 and 23, respectively.

Second image is *Angelina* (Fig. 27), which is a color Bitmap File Format (lossless) of 1348-by-1078 pixels with 24 bit-per-pixel (bpp).

We have the same noise as in the previous case.

Fig. 27 (top-left) shows us the original image used in this experiment; noisy image (top-right); the filtered images, processed by using classical directional smoothing (middle-left), and quantum directional smoothing techniques (middle-right), respectively. Besides, Fig. 27 (down-center) shows the difference pixel-to-pixel between classical and quantum versions. As we can see, there are values of pixels where the difference between two versions is remarkably sensitive here too. However, such is less than in the previous case. It has to do with a lower edges richness of *Angelina* vs *Agus in Miami*. Another important responsible factors for this difference are constituted by: a) *Agus in Miami* has higher values in its LUMA [14-17], b) *Agus in Miami* has more brightness and contrast; and, c) *Agus in Miami* is larger than *Angelina*.

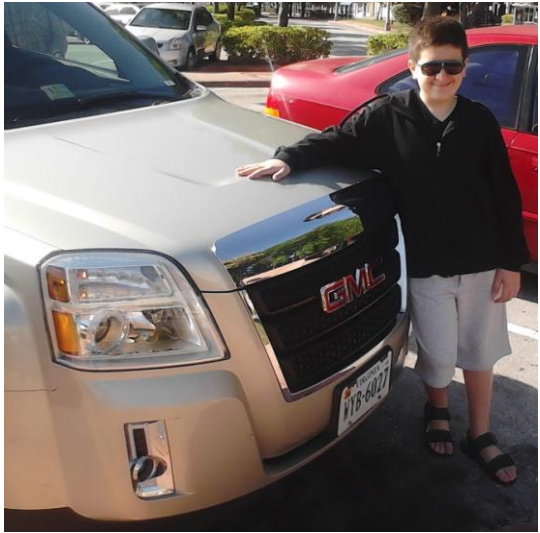
Original

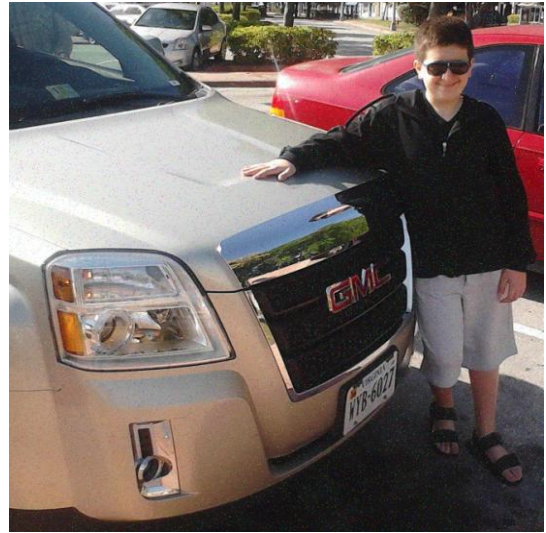
Noisy

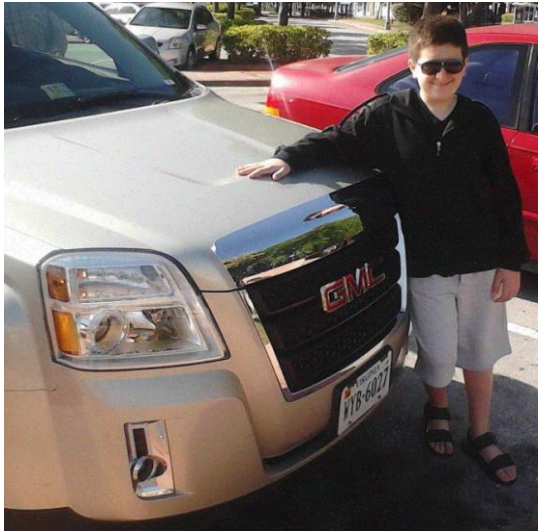
Classical Directional Smoothing

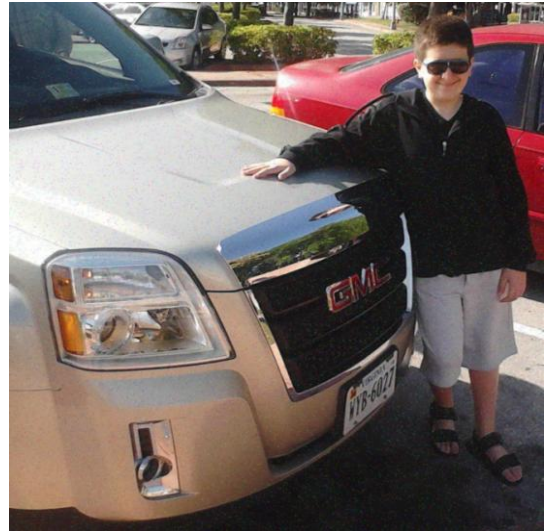
Quantum Directional Smoothing

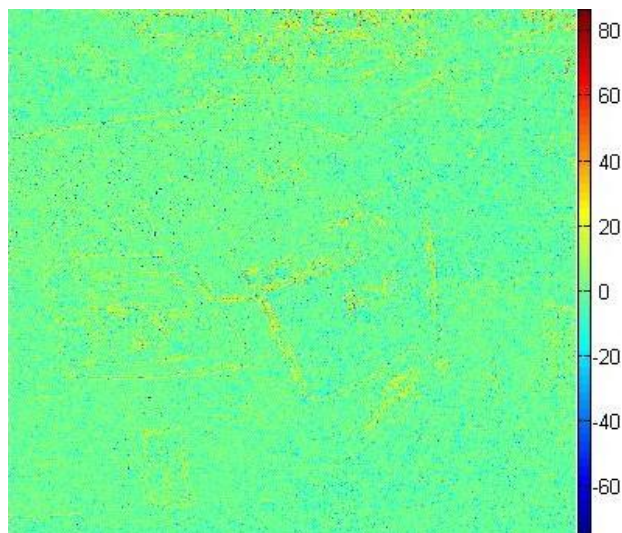
Error pixel-to-pixel

**Fig. 26** Denoising for Agus in Miami.

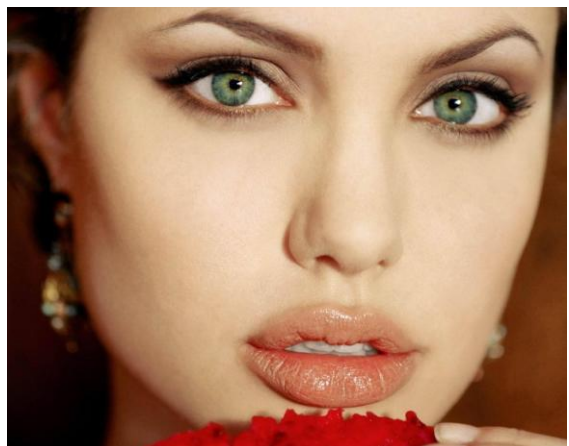
Original

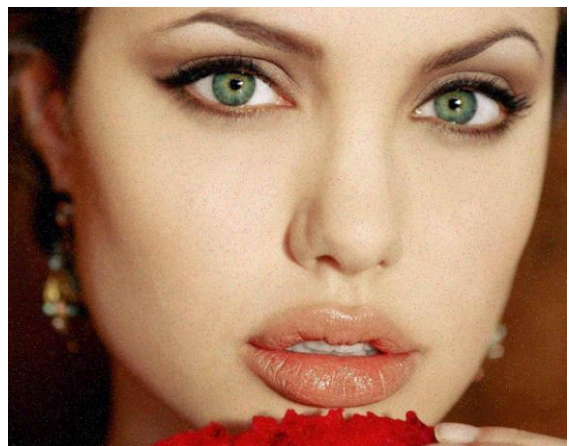
Noisy

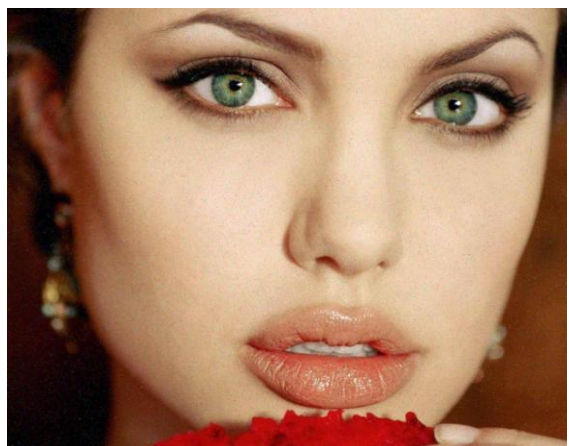
Classical Directional Smoothing

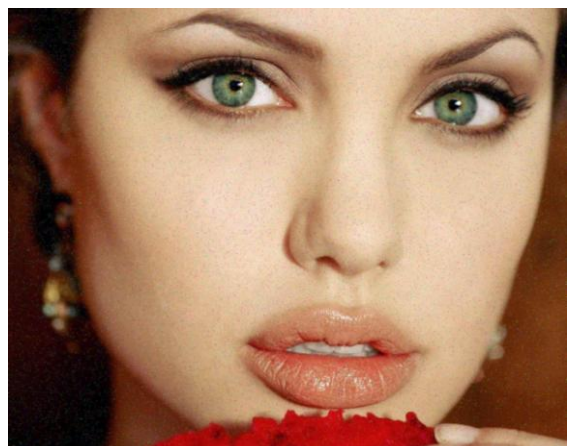
Quantum Directional Smoothing

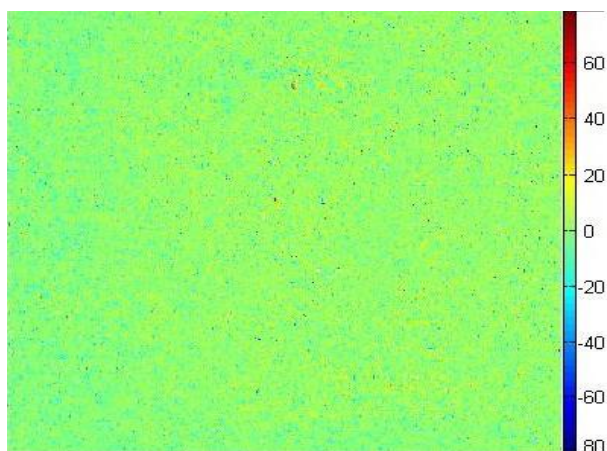
Error pixel-to-pixel

**Fig. 27** Denoising for Angelina.

This later attribute seems irrelevant to naked eye, however, it is not, since, to process more qubits, it automatically increases the detrimental intervention of bad (or poorly) modeled noise. Besides, a larger image means more openness in the time window of the process which may be more exposed to quantum decoherence [22, 118-123]. This is a topic that should be further investigated if we want to process images of very high resolution in a quantum computer. In Table V, we can see that PCQR is bigger than last case, however, it is a low value too.

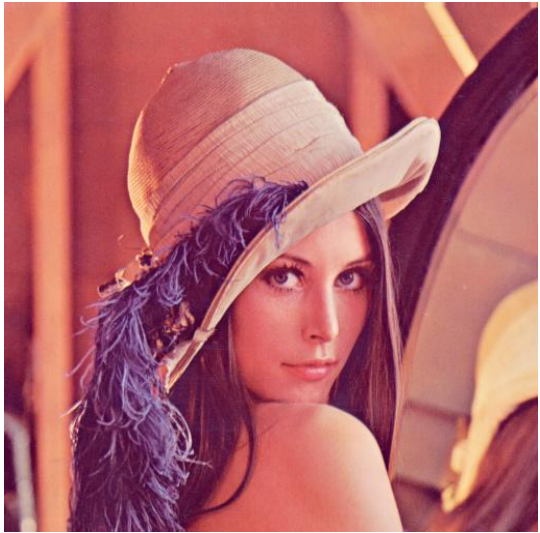
Original

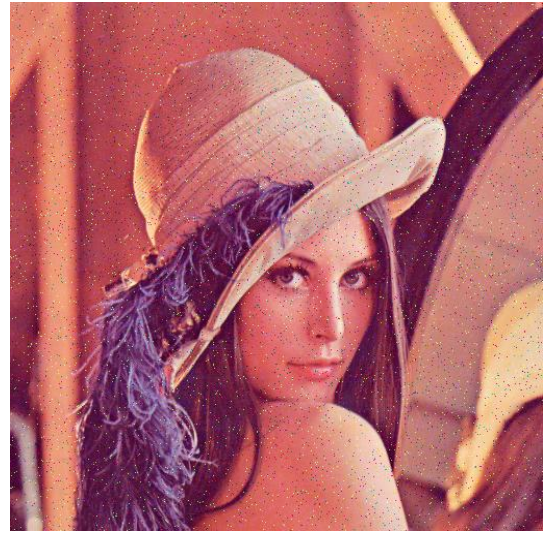
Noisy

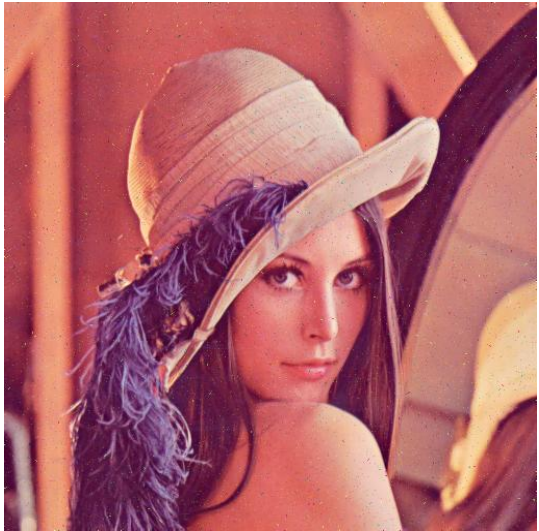
Classical Directional Smoothing

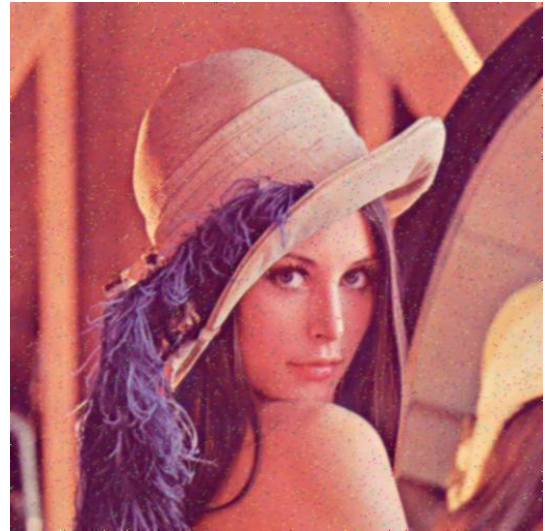
Quantum Directional Smoothing

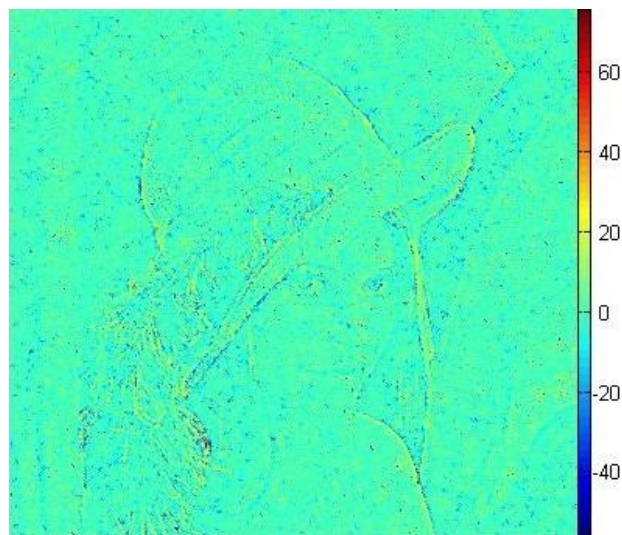
Error pixel-to-pixel

**Fig. 28** Denoising for Lena.

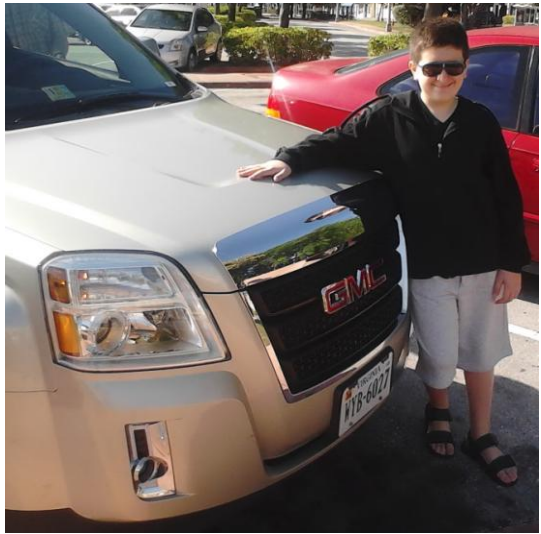
Original

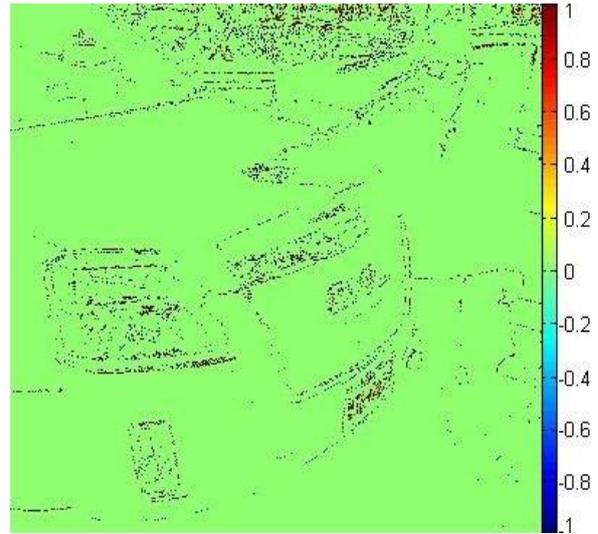
Error pixel-to-pixel

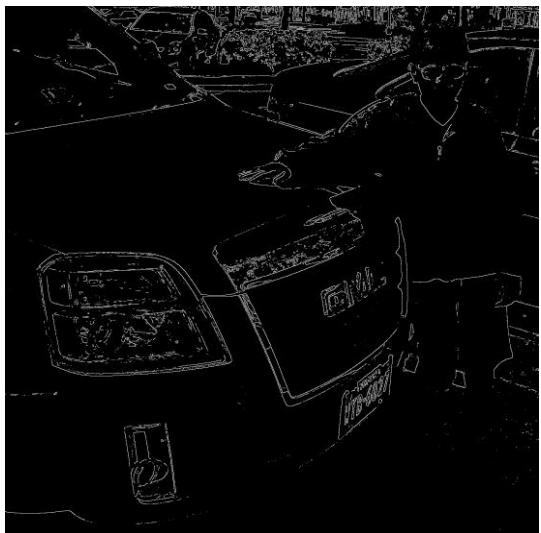
Classical Sobel

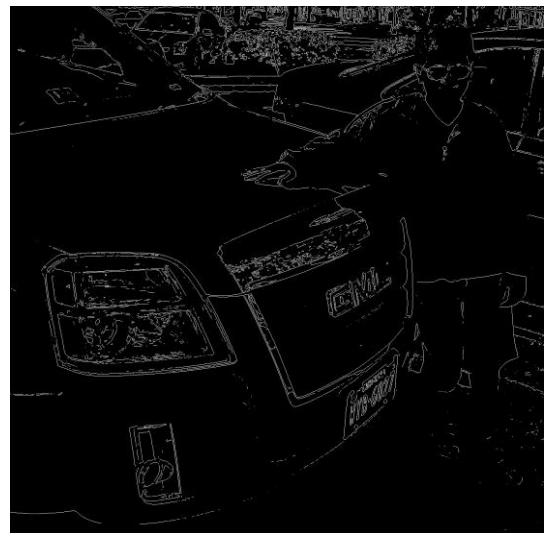
Quantum Sobel

**Fig. 29** Edge Detection for Agus in Miami.

Finally, last image is *Lena* (Fig. 28), which is a color Bitmap File Format (lossless) of 512-by-512 pixels with 24 bit-per-pixel (bpp).

In this case, identical considerations to previous cases are used regardig to present noise in the image. Tests with other types of noise gave identical comparative results

Fig. 28 (top-left) shows us the original image used in this experiment; noisy image (top-right); the filtered images, processed by using classical (middle-left), and quantum directional smoothing techniques (middle-right), respectively. Besides, Fig. 28 (down-center) shows the difference pixel-to-pixel between classical and quantum versions of filtering process.

Based on the analysis of the comparison between *Agus* and *Angelina*, we can understand why this picture is showing the best fit between classical and quantum version of filters. This can be seen clearly in the metrics of Table V, where we obtain the lower values of MAD and MSD from three images so far treated. In return, this image has the highest value of PCQR.

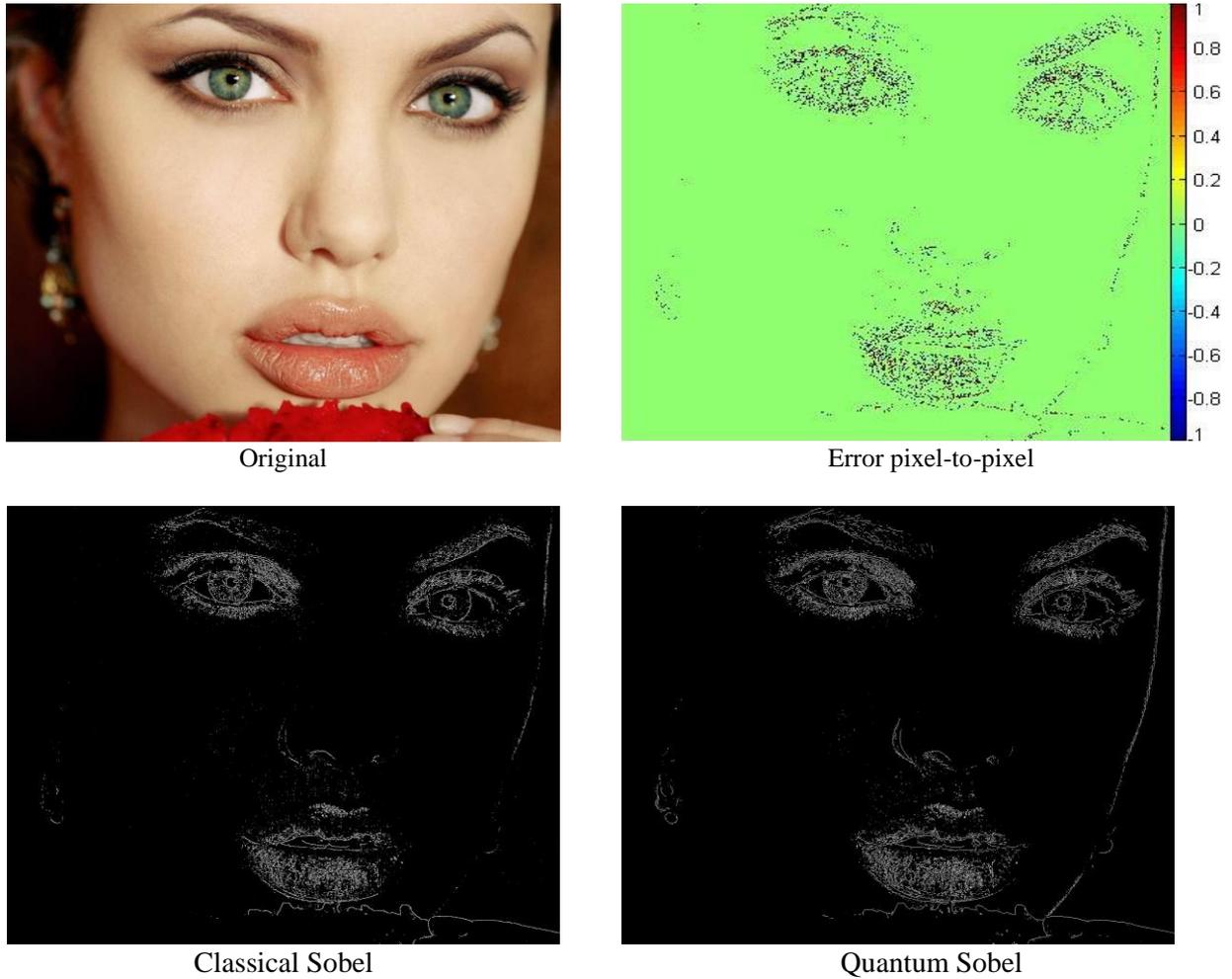

| | |
|---|---|
| Original | Error pixel-to-pixel |
| Classical Sobel | Quantum Sobel |

**Fig. 30** Edge Detection for Angelina.

*Edge-detection of multimedia images*

For these experiments, all images are subjected to a built-in MATLAB® function called *rgb2gray*. This function converts a color image into gray [143].

In Fig.29, we can see original image of *Agus in Miami* (top-left), while in the same image (down-left) we have edge-detection via classical version of Sobel filtering, (down-right) we have quantum version of Sobel filtering. Besides, Fig. 29 (top-right) shows the difference pixel-to-pixel between classical and quantum versions of filtering process. These differences are significant.

TABLE VI
METRICS OF EDGE DETECTION: CLASSICAL VS. QUANTUM

| IMAGE | MAD | MSD | PCQR |
|---|---|---|---|
| AGUS | 0.0151 | 0.0151 | 18.2006 |
| ANGELINA | 0.0110 | 0.0110 | 19.5681 |
| LENA | 0.0169 | 0.0169 | 17.7322 |

Table VI show us the metrics for this experiment. Notice that MAD and MSD are significantly better than in despeckling and denoising cases. However, PCQR is lousy. This is because the maximum value of the classic image is not 255 (as shown in Equation 75) but one. Remember that, the images to enter in Equation (75) are not the original color image or gray but the resulting of Sobel filter.

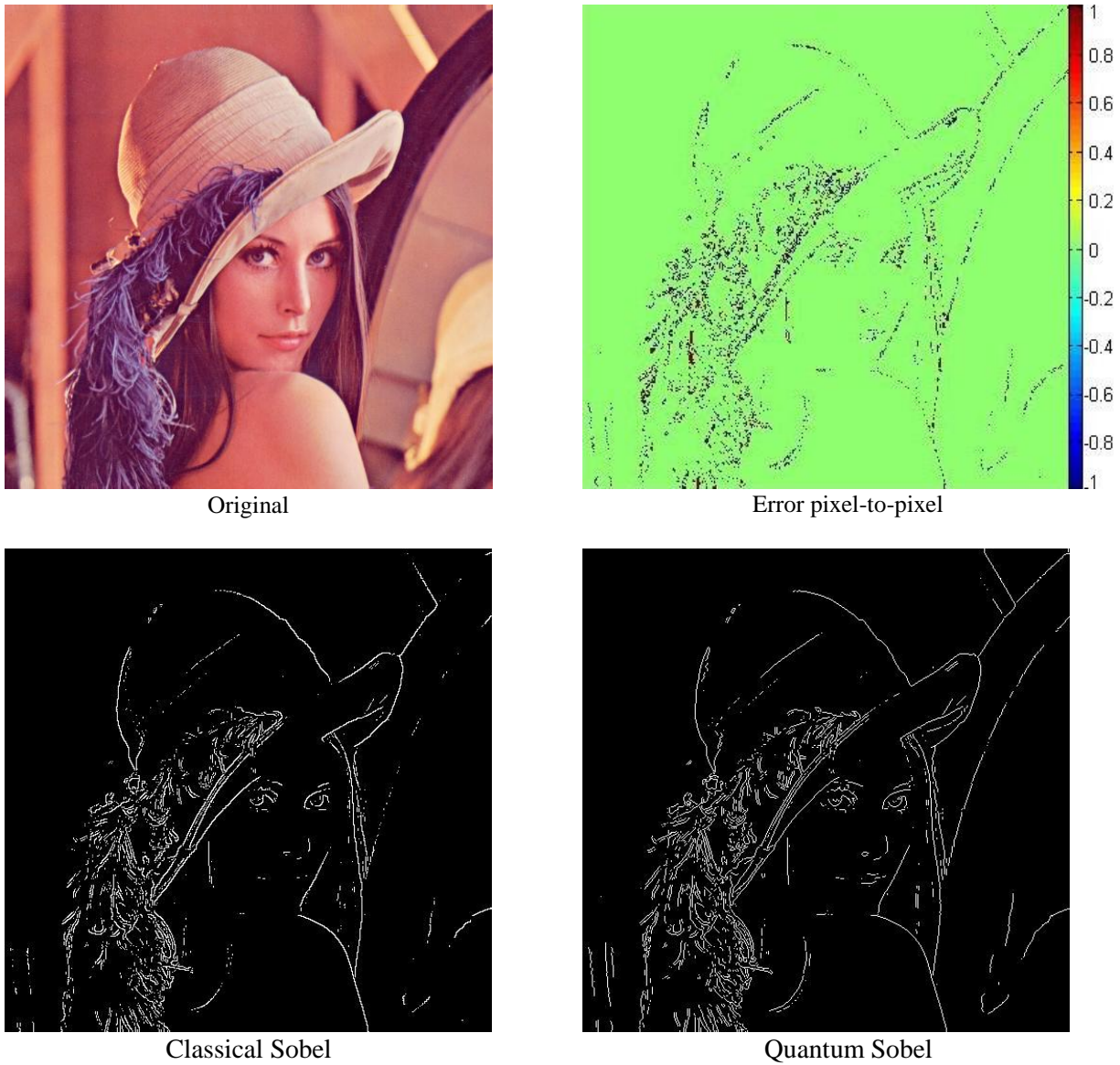

**Fig. 31** Edge Detection for Lena.

In Fig.30, we can see original image of *Angelina* (top-left), while in the same image (down-left) we have edge-detection via classical version of Sobel filtering, (down-right) we have quantum version of Sobel filtering. Besides, Fig. 30 (top-right) shows the difference pixel-to-pixel between classical and quantum versions of filtering process. Newly, here too, these differences are significant.

In Table VI, we can see that the metrics are much better than in the previous case. The most important responsible factors for this difference are constituted by: a) *Agus in Miami* has higher values in its LUMA [14-17], b) *Agus in Miami* has more brightness and contrast; and, c) *Agus in Miami* is larger than *Angelina*. That is to say, the reasons why Angelina gives better results than Agus are coincident for filtering and edge detection.

Finally, newly, and as it was predictable, Lena (Fig. 31) has the worst results of the three.

In this section we present a series of experiments. They showed the gap between classical and quantum versions of the filters due to state and measurement noise.

# 9 Conclusions and Future Works

In this paper we have presented -in order- the following advances:

## 9.1 Pole-to-pole Axis Only (PAO)

Basically, PAO consists of a new criterion, Logic, and Arithmetic based on projections onto vertical axis of Bloch's Sphere exclusively. This approach allowed us:

1) a simpler development of logic and arithmetic quantum operations, where they will closer to those used in the classical digital image processing algorithms,
2) incorporation of quantum converter before quantum algorithm (circuit or gate), or classical converters after measurement of vertical projection for each qubit, which allow develop quantum arithmetic operations (on a Hermitian context) as if it were in a classical context in the real field
3) building simple and robust classical-to-quantum and quantum-to-classical interfaces, based on the need to measure only the projections on the vertical axis (i.e., *z*-axis, or pole-to-pole axis)
4) build a simple and robust optimal estimator of quantum states avoiding Kroeneker's Algebra

## 9.2 Optimal State Estimator (OSE)

An optimal estimator of quantum states based on a modified Kalman's Filter was presented in this work. Such estimator acts after state measurement, allowing to obtain an optimal estimation of quantum state resulting to the output of any quantum image algorithm (circuit or gate). Although the OSE allows us a complete estimation of the quantum state, however, we are interested in accurate measurement and estimation of the vertical component only, thanks to PAO, which dramatically simplifies its equations.

## 9.3 Classical-to-quantum and quantum-to-classical interfaces

We developed two modes of quantum-to-classical interface and one mode of classical-to-quantum interface according to the above (i.e., based on PAO). In addition to everything mentioned in the corresponding section, once obtained the $\mu$ based on its $\alpha$, to reach the levels of the external image (classical), just need an equalizer and a rounder. These interfaces represent the purest functional interpretation on how they should work the same. This is the greatest contribution of PAO.

## 9.4 Three new metrics

In a special section on metrics and simulations, three new metrics based on the comparison between the classical and quantum versions algorithms for filtering and edge detection of images were presented. Notable differences between the results of classical and quantum versions of such algorithms (outside and inside of quantum computer, respectively) showed the need for modeling state and measurement noise inside estimation scheme.

Summing-up, our quantum image processing filter and detects edges, however, it is far from the classical image processing due to our inability to model appropriately the intervening noises. The later is one of the remaining tasks. The other is to apply the developed in this work to any quantum algorithm and Quantum Physics in general.

Finally, both classical techniques (i.e., denosing/despeckling and edge-detection) were implemented in MATLAB® R2014a (Mathworks, Natick, MA) [143] on a notebook with Intel® Core(TM) i5 CPU M 430 @ 2.27 GHz and 6 GB RAM on Microsoft® Windows 7© Home Premium 32 bits. Besides, a simulated version of quantum implementations were done on a GPU cluster, NVIDIA® Tesla© 2050 GPU [144] with a peak performance of approximately 500 GFLOPS, with an achieved performance of approximately 250

GFLOPS in OpenCL. The GPU needed approximately 2.5 GB of bandwidth with InfiniBand connectivity at quad data rate (QDR) QLogic® [145] or 40 Gb speeds.

**Acknowledgments** M. Mastriani thanks Prof. Dr. Salvador E. Venegas-Andraca, Assistant Professor of Mathematics and Computer Science from Monterrey Institute of Technology-State of Mexico Campus for being the creator of this discipline, by share it -so selflessly- with all humanity, and for his tremendous help and support. Besides, I wish to thank all the technical staff of the various laboratories of the National Commission of Atomic Energy for the help they gave me in the preparation of experiments. It is impossible to name them all here, simply, thank you all for all.